\documentclass{aa}  
\usepackage{graphicx}
\usepackage{txfonts}
\usepackage{multirow}
\usepackage{xcolor,colortbl}
\usepackage{subcaption}
\usepackage{adjustbox}

\begin{document}

   \title{Black hole and host galaxy growth in an isolated $z\sim 6$ QSO observed with ALMA}

   \author{R. Tripodi
    \inst{1,2,3}\thanks{\email{roberta.tripodi@inaf.it}}
    \and C. Feruglio
    \inst{2,3}
    \and F. Fiore 
    \inst{1,2,3}
     \and M. Bischetti
    \inst{2}
     \and V. D'Odorico
    \inst{2,3,4}
   \and S. Carniani
   \inst{4}
   \and S. Cristiani
   \inst{2,3,5}
   \and S. Gallerani
   \inst{4}
   \and R. Maiolino
   \inst{6,7,8}
   \and A. Marconi
   \inst{9,10}
   \and A. Pallottini
   \inst{4}
   \and E. Piconcelli
   \inst{11}
   \and L. Vallini
   \inst{4}
   \and T. Zana
   \inst{4}
          }

   \institute{Dipartimento di Fisica, Università di Trieste, Sezione di Astronomia, Via G.B. Tiepolo 11, I-34131 Trieste, Italy
        \and
        INAF - Osservatorio Astronomico di Trieste, Via G. Tiepolo 11, I-34143 Trieste, Italy
         \and
         IFPU - Institute for Fundamental Physics of the Universe, via Beirut 2, I-34151 Trieste, Italy
         \and
         Scuola Normale Superiore, Piazza dei Cavalieri 7 I-56126 Pisa, Italy
         \and
         INFN - National Institute for Nuclear Physics, via Valerio 2, I-34127 Trieste, Italy
         \and
         Institute of Astronomy, University of Cambridge, Madingley Road, Cambridge CB3 0HA, UK
         \and
         Kavli Institute for Cosmology, University of Cambridge, Madingley Road, Cambridge CB3 0HA, UK
         \and 
         Department of Physics and Astronomy, University College London, Gower Street, London WC1E 6BT, UK
         \and
         Dipartimento di Fisica e Astronomia, Università di Firenze, via G. Sansone 1, Sesto F.no (Firenze), Italy
         \and
         INAF - Osservatorio Astrofisico di Arcetri, Largo E. Fermi 5, 50125, Firenze, Italy
         \and
         INAF - Osservatorio Astronomico di Roma, Via Frascati 33, I-00040 Monte Porzio Catone, Italy
         \\
             }

   \date{Accepted July 6, 2022}

  \abstract
{ The outstanding mass growth of supermassive black holes (SMBHs) at the epoch of reionisation and its relation to the concurrent growth of their host galaxies poses challenges to theoretical models aimed at explaining how these systems formed on short timescales (<1 Gyr). To trace the average evolutionary paths of quasi-stellar objects (QSOs) and their host galaxies in the plane of BH mass to host mass ($M_{\rm dyn}$), we compare the star formation rate (SFR), derived from the accurate estimate of the dust temperature and the dust mass ($T_{\rm dust}, M_{\rm dust}$) based on infrared and sub-millimeter  (sub-mm) spectral energy distribution (SED), with the BH accretion rate, derived from $L_{\rm bol}$ based on X-ray and optical and ultraviolet SED. To this aim, we analysed a deep ALMA observation of the sub-mm continuum, [CII], and H$_2$O of the $z\sim 6$ QSO J2310+1855 with a resolution of $900$ pc, which enabled a detailed study of dust properties and cold gas kinematics. 
    We performed an accurate SED analysis obtaining a dust temperature of $T_{\rm dust} = 71\pm 4$ K, dust mass $M_{\rm dust}= (4.4 \pm 0.7) \times 10^8\ \rm M_{\odot}$ , and total far-infrared luminosity of $L_{\rm TIR} = 2.5^{+0.6}_{-0.5} \times 10^{13}\ \rm L_{\odot}$. The implied active galactic nuclei (AGN) - corrected ${\rm SFR} = 1240^{+310}_{-260}\ \rm M_{\odot}yr^{-1}$ is a factor of 2 lower than previously reported for this QSO. We measured a  gas-to-dust ratio of GDR$= 101\pm 20$. The dust continuum and [CII] surface brightness profiles are spatially extended out to $r\sim6.7$ kpc and $r\sim5$ kpc, respectively,  with half-light radii of  0.9 and 1.1 kpc for the dust and gas, respectively. The derived gas surface density, $\Sigma_{\rm gas}$, and star formation rate density, $\Sigma_{\rm SFR} $, place the J2310+1855 host galaxy above the Kennicutt-Schmidt relation. We derived a best estimate of the dynamical mass $M_{\rm dyn} = 5.2\times 10^{10}\ \rm M_{\odot}$ within $r = 1.7$ kpc based on a dynamical model of the system with a rotating disk inclined at $i = 25$ deg. The Toomre parameter profile across the disk is $Q_{\rm gas}\sim 3$ and implies that the disk is unstable.  We found that ${\rm SFR}/M_{\rm dyn}>\dot M_{\rm BH}/M_{\rm BH}$, suggesting that AGN feedback might be efficiently acting to slow down the SMBH accretion, while stellar mass assembly is still vigorously taking place in the host galaxy.
     In addition, we were also able to detect high-velocity emission on the red and blue sides of the [CII] emission line that is not consistent with disk rotation and traces a gaseous outflow. We derived an outflow mass $M_{\rm out} = 3.5\times 10^8\ \rm M_{\odot}$, and a mass outflow rate in the range $\dot M_{\rm out}= 1800-4500\rm ~M_\odot yr^{-1}$. The implied $\dot E_{\rm out} \sim 0.0005-0.001\ L_{\rm bol}$ is in agreement with the values observed for ionised winds. 
     For the first time, we mapped a spatially resolved water vapour disk through the H$_2$O v=0 $3_{(2,2)}-3_{(1,3)}$ emission line detected at $\nu_{\rm obs} = 274.074$ GHz, whose kinematic properties and size are broadly consistent with those of the [CII] disk. The luminosity ratio $L_{\rm H_2O}/L_{\rm TIR}= 1.4\times 10^{-5}$ is consistent with line excitation by dust-reprocessed star formation in the interstellar medium of the host galaxy. 
    }

   \maketitle


\section{Introduction}
\label{sec:intro}

Luminous quasi-stellar objects (QSOs), powered by accretion onto supermassive black holes (SMBHs), already exist at the epoch of reionisation, when the Universe was only 0.5-1 Gyr old. Their BH masses are not lower than those of hyper-luminous QSOs at lower redshift, meaning that BH growth had to be a fast process, and that the process had to stop with a similar high efficiency after the rapid build-up. How huge BHs formed and grew in such a short time is indeed highly debated \citep{volonteri2010,johnson2016}, particularly as they lie above the local $M_{\rm BH}$ - $M_{\rm dyn}$ correlation and thus follow the BH-dominance growth path \citep{volonteri2012}. 
Once started, it is unknown what slowed the BH growth down and when this process occurred, leading towards the symbiotic growth with the host galaxy observed in the local Universe. Candidate processes are inefficient gas accretion and/or feedback through BH winds. At the same time, the host galaxies of high-z QSOs are likely growing rapidly. Therefore, the onset of significant BH feedback hampering BH growth would mark the transition from  a phase of BH dominance to a phase of symbiotic growth of the BH and the galaxy. Cosmological hydrodynamic simulations of early BH and galaxy evolution support this scenario by identifying z$\sim$6-7 as the transition epoch during which QSO feedback increases in strength and starts to significantly slow down BH growth. Moreover, the QSO host galaxies provide a unique opportunity to characterize both the physical properties of the interstellar medium (ISM) in such extreme conditions (e.g. \citealt{bertoldi2003a, bertoldi2003b,decarli2018,venemans2020,neeleman2021, Pensabene2021}) and to study the formation and build-up of massive galaxies in the early Universe in detail. 

In the past few decades, \textit{Herschel}, the Northern Extended Millimeter Array (NOEMA), the Very Large Array (VLA), and particularly the Atacama Large Millimeter/sub-millimeter  Array (ALMA) have been able to probe and give insights into the properties of the gas and dust inside the QSO host galaxies, allowing us to derive the dynamical masses, star formation activity, and ISM properties. Observations using, for instance, the instruments on \textit{Herschel}, NOEMA, and ALMA have detected the dust continuum in the host galaxies of many $z \sim 6$ QSOs, with far-IR (FIR) luminosities of $10^{11-13}\ \rm L_{\odot}$ and dust masses of about $10^{7-9}\ \rm M_{\odot}$ \citep{decarli2018, carniani+19, shao2019}. The rest-frame FIR continuum emission in these sources originates from dust heated by the ultraviolet (UV) radiation from young and massive stars in the host galaxies and the QSO radiation field. It is often hard to determine the temperature and mass of the dust precisely since they are both highly degenerate and the FIR spectral energy distribution (SED) is sparsely probed,  often relying on single-frequency continuum detection. However, if multi-frequency ALMA observations are available in the FIR, it is possible to constrain the dust temperature and mass with statistical uncertainties <10\% (see e.g. \citealt{carniani+19}), implying a high accuracy in the determination of the star formation rate (SFR). An accurate estimate of the dust mass would also allow us to derive the molecular gas mass of the host galaxy through the gas-to-dust ratio (GDR). Although it is possible to directly probe the molecular reservoirs of the QSO host galaxies using the rotational transitions of the carbon monoxide (CO; e.g. \citealt{vallini2018, madden2020}), very few high-z QSOs are observed in CO because this emission line is typically very faint at high z. The GDR indeed has often been assumed in order to compute the gas mass, implying an high degree of uncertainty in its estimate. Studies of $z\sim 2.4-4.7$ hyper-luminous QSOs show that the GDR spans a broad range of values, [100-300], with an average GDR $\sim 180$ \citep{bischetti2021}, consistent with the results found for sub-millimetre galaxies out to $z\sim3-5$ with GDR$\sim150-250$ (e.g. \citealt{saintonge2013,miettinen2017}). In low-z galaxies, a GDR$\sim 100$ is typically observed \citep{draine2007, leroy2011}, implying that the GDR increases with redshift. However, if we are able to derive a reliable estimate of the gas mass from CO, this could be used, together with the accurate estimate for the dust mass, to determine the GDR instead of assuming it, and to use it for other high-z QSOs. 

The ISM of the QSO host galaxies has compact sizes of a few kiloparsec (e.g. \citealt{wang2013, shao2017, venemans2020, neeleman2021}) and shows massive gas reservoirs \citep{feruglio2018}. The [CII] $\lambda$158$\mu$m is the dominant cooling line of the ISM and the brightest emission line, almost unaffected by attenuation. Therefore, it is the preferred tracer for studying the ISM and provides valuable information about cold, warm neutral and mildly ionised ISM \citep{cormier2015,olsen2018}. Its 158$\mu$m transition predominantly arises from photodissociation regions (PDRs; \citealt{hollenbach1999}) at the interface of the atomic and molecular media in the outskirts of molecular clouds in galaxy star-forming regions. 
Through kinematical studies of [CII], dynamical masses have been derived of $<10^{11}\ \rm M_{\odot}$ (e.g. \citealt{shao2017, Pensabene2021, izumi2021a, neeleman2021}), which place most of the $z\sim 6$ QSOs above the\ $M_{\rm BH}-M_{\rm dyn}$ relation by a factor of $\sim 3-10$ . This in turn allows us to distinguish among the different growth paths of the SMBHs and their host galaxies (BH dominance, symbiotic growth, or BH adjustment, see \citealt{volonteri2012}).

In the past decade, the strong coupling between the ISM (and also the circum-galactic medium, CGM) and the SMBH energy output has been observed and was modelled theoretically. It occurs as mechanical and radiative QSO-driven feedback processes, and it affects the evolution of the whole galaxy. In particular, very powerful mechanisms that efficiently deposit energy and momentum into the surrounding gas are found to be QSO-driven outflows \citep{faucher2012, zubovas2012}. They have been extensively studied and detected from the local Universe back to the epoch of reionisation in all gas phases, at all spacial scales, from sub-parsec to several kiloparsecs, even with high kinetic power (up to a few percent of the bolometric luminosity) and with mass outflow rates exceeding the star formation rate \citep{feruglio2010, maiolino2012, cicone2015, aalto2015, fiore2017, bischetti2019b}. Because of observational limitations, we unfortunately have very few detections of outflows in high-z QSOs \citep{maiolino2005,bischetti2019,izumi2021a,izumi2021b}, but a precise determination of their masses, extension, and occurrence would give us valuable insights into the onset of the active galactic nucleus (AGN) feedback in the first QSOs. 

Finally, the broad-band coverage of ALMA allows the serendipitous detection of additional emission lines arising from the galaxy ISM, such as $\rm H_2O$, which can be used to provide additional constraints on the ISM properties. Water vapour emission lines have been detected from $z>3$ galaxies and QSOs (e.g. \citealt{vanderwerf2011, combes2012, omont2013, reichers2013}), but they are still rare and unresolved, if detected, at $z\gtrsim 6$ (e.g. \citealt{Yang2019, Pensabene2021} and references therein). Although little information is still available at high-z, a correlation between $L_{\rm H_2O}$ and the total infrared luminosity has been found and studied, suggesting that the water vapour lines would be excited by an IR-pumping mechanism from the ISM UV radiation field \citep{Yang2019, Pensabene2021}.

We present new high-resolution ALMA observation of the [CII], the H$_2$O emission lines, and the sub-millimeter (sub-mm) continuum of QSO SDSS J231038.88+185519.7 (hereafter J2310+1855 or J2310). J2310, first discovered in SDSS \citep{jiang2006,wang2013}, is one of the most FIR-luminous QSOs and one of the brightest optical QSOs known at $z\sim 6$, with $L_{\rm bol}=9.3 \times 10^{13}\ \rm L_\odot$. The redshift measured with the QSO rest-frame UV line emission is $z = 6.00\pm 0.03$ \citep{wang2013}. \citet{feruglio2018} detected and analysed the CO(6-5) and [CII] emission lines and the sub-millimetre continuum of J2310, deriving a size of the dense molecular gas of $2.9\pm 0.5$ kpc and of $1.4 \pm 0.2$ kpc for the 91.5 GHz dust continuum and a molecular gas mass of $M({\rm H}_2)=(3.2\pm 0.2)\times 10^{10}\rm M_{\odot}$. They estimated a dynamical mass of $M_{\rm dyn} =  (4.1^{+9.5}_{-0.5})\times 10^{10}\rm M_\odot$, measuring a disk inclination of $i\sim50$ deg. They also inferred the BH mass from the CIV emission line, measured in the X-shooter/VLT spectrum of the QSO, obtaining $M_{\rm BH}=(1.8\pm 0.5)\times 10^9\rm M_\odot$. Recently, \citet{shao2019} presented a detailed analysis of the FIR and sub-mm SED and derived a dust temperature of $T\sim 40$ K, a dust mass of $M_{\rm dust}= 1.6\times 10^9\rm M_\odot$, a FIR luminosity $L_{\rm FIR}^{8-1000 \mu m}=1.6\times  10^{13}\ \rm L_{\odot}$, and an SFR$= 2400-2700\ \rm M_\odot yr^{-1}$. \citet{dodorico2018} detected a very metal-poor, proximate damped Lyman $\alpha$ system (DLA) located at z=$5.938646 \pm 0.000007$ in the X-shooter/VLT spectrum of J2310, which was associated with a CO emitting source at $z = 5.939$. This source, called Serenity-18, was detected through its CO(6-5) emission line at [RA, DEC] =  23:10:38.44, 18:55:21.95.  

\begin{table*}
\caption{Summary of the ALMA observations and their properties }    
\centering       
\label{table:obs}                        
\begin{tabular}{c c c c c c c c}        
\hline\hline  
Dataset & Project ID & RA, DEC & Central Freq. & Baselines & Synth. beam& R.m.s. cont. & R.m.s. cube  \\
& & (J2000) & (GHz) & (m) & (arcsec$^2$) & ($\mu$Jy/beam) & (mJy/beam) \\ 
\hline
I & 2019.1.00661.S &  23:10:38.44, 18:55:21.95 & 264.695  & $15-2517$ & 0.26$\times$0.21$^a$  &   8.8 & 0.23$^c$  \\
  &                &                        &             &           &
  0.17$\times$0.15$^b$  & 9.1   &  0.23\\
II & 2019.1.01721.S &  23:10:38.88, 18:55:19.70 & 265.54 & $15-313$ &  1.6$\times$1.3 & 56 & 0.4$^c$ \\ 
\hline
                                 
\end{tabular}
 \flushleft 
\footnotesize {{\bf Notes.} (a) Natural weighting. (b) Briggs weighting with robust=0.5.  (c) Per 8.5 $\rm km\ s^{-1}$ spectral channel. Dataset I is a new ALMA observation that is analysed in this paper for the first time. Dataset II is an archival ALMA observation without published results.}
\end{table*}

 The paper is organised as follows. The observations are described in Sect. \ref{sec:obs}; in Sect. \ref{sec:res} we show the results for the continuum emission, the [CII], and the H$_2$O v=0 $3_{(2,2)}-3_{(1,3)}$ emission lines. In Sect. \ref{sec:disc} we report on the analysis of the SED of J2310+1855, the [CII], H$_2$O distributions and kinematics, and the environment of the QSO. A discussion and summary are presented in Sect. \ref{sec:summary}, and the conclusions are reported in Sect. \ref{sec:concl}.  

We adopted a $\Lambda$CDM cosmology from \citet{planck2016}: $H_0=67.7\ \rm km\ s^{-1}\ Mpc^{-1}$, $\Omega_m = 0.308$ and $\Omega_{\Lambda} = 0.7$. Thus, the angular scale is $5.84$ kpc/arcsec at $z=6$.

\begin{figure}
        \centering
        \includegraphics[width=1\linewidth]{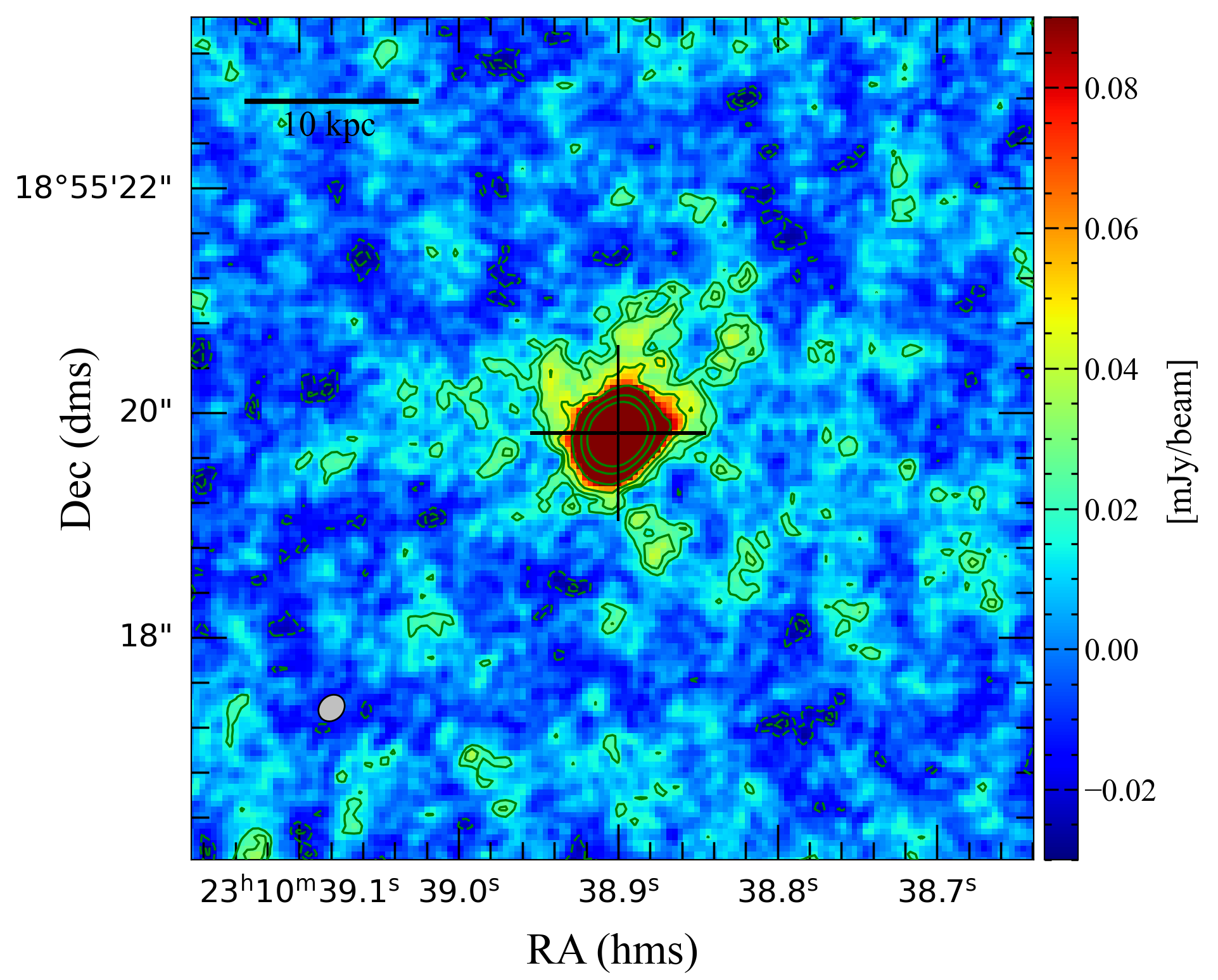}
        \caption{260 GHz dust continuum map of QSO J2310+1855 (levels $-4,-3,-2,2,3,5,10,25,\text{and }50\sigma$, $\sigma = 9.15\ \mu$Jy/beam). The clean beam ($0.26\times 0.21\rm \ arcsec^2$) is indicated in the lower left corner of the diagram. The cross indicates the position of the continuum peak.}
        \label{contj2310}
\end{figure}

\section{Observations}
\label{sec:obs}
We analysed the dataset 2019.1.00661.S from the ALMA 12m array towards the QSO SDSS J2310+1855. This observation was pointed towards [RA, DEC] =  23:10:38.44, 18:55:21.95 with a central frequency of 264.695 GHz. The primary beam of the observation includes QSO J2310+1855, the frequency setup covers the [CII] $^2P_{3/2} - ^2P_{1/2}$ emission line from the QSO and the adjacent continuum, and the [CII] emission from a DLA located on the QSO sightline at $z=5.938646$. This observation, with a total integration time of 4.3 hours and a maximum baseline of 2517 m, was primarily designed to detect [CII] emission towards Serenity-18 \citep{dodorico2018}, at the same time allowing high-resolution imaging of QSO J2310.

The calibration of visibilities was performed for all datasets through the Common Astronomy Software Applications pipeline (CASA; \citealt{mcmullin2007}), version 5.1.1-5. 
To image the QSO, we applied  \texttt{phasecenter} within \texttt{tclean} to place the QSO at phase tracking centre. 
We imaged the data using both natural and Briggs weighting with a robust parameter equal to 0.5, the latter in order to maximise the angular resolution, and we applied a $3\sigma$ cleaning threshold. 
We imaged the continuum by collapsing all line-free channels\footnote{Lines were detected in channel ranges $107\sim185$ for spw0 ($\rm H_2O$) and $72\sim168$ ([CII]) for spw1.}, selected by inspecting the visibilities in all spectral windows. We used the CASA task uvcontsub to fit the continuum
visibilities in the line-free channels and obtained continuum-subtracted cubes with spectral channels of width 8.5 $\rm km\ s^{-1}$. 
To obtain continuum-subtracted data cubes, we fitted the continuum in upper and lower side band (USB and LSB) with a first-order polynomial, since the continuum shows a non-negligible slope. 
The clean beams we obtained are ($0.26 \times 0.21$) arcsec$^2$ for natural and ($0.17 \times 0.15$ ) arcsec$^2$ for Briggs weighting. The maximum spatial resolution we reached is about 0.9 kpc at the rest frame of the QSO and is obtained in the Briggs data cube. The r.m.s. noise reached is 8.8 $\mu$Jy/beam in the continuum and 0.23 mJy/beam per 8.5 $\rm km\ s^{-1}$ channel for the natural weighted maps (Table \ref{table:obs}). The high resolution and sensitivity of this observation enabled us to perform a detailed analysis of the [CII] disk through dynamical modelling, which was not possible before. We retrieved a new precise estimate for the dynamical mass (see Sect. \ref{sec:ciikinematics}). Moreover, we were able to spatially resolve the water vapour emission detected at $\sim 274$ GHz (see Sect. \ref{sec:h2oemission}) for the first time.

We also analysed an additional dataset (II) from the ALMA archive (project 2019.1.01721.S). This has a maximum baseline of 313 m and central frequency of 365.54 GHz, and it also covers the [CII] emission line from both QSO J2310+1855 and the DLA at $z=5.938646$.
We imaged dataset II  using the same method as described above, and applied only natural weighting. We obtained a clean beam of (1.6 x 1.3) arcsec$^2$. The r.m.s. noise is 56 $\mu$Jy/beam in the continuum and 0.4 mJy/beam per 8.5 $\rm km\ s^{-1}$ channel in the data cube.  Combination of the two datasets was performed but did not produce gain in sensitivity or image quality because the angular resolutions of the two datasets are very different. We therefore did not use it in the following analysis.

We analysed the datasets, centred on J2310+1855, present in the ALMA archive and more recent than 2011 in order to derive the continuum flux densities at different frequencies for QSO J2310+1855. For all datasets, calibration and imaging were performed as outlined above, and natural weighting was applied everywhere with a detection threshold of 3$\sigma$.

\begin{table*}
\caption{Measurements and derived quantities for the emission lines and dust continuum in SDSS J231038.88+185519.7.}
\centering       
\label{table:misure}                        
\begin{tabular}{l l c c | c }        
\hline\hline  
& & \multicolumn{2}{c|}{Emission Lines} & \multirow{2}{7em}{260 GHz Continuum} \\
& & CII $^2P_{3/2} - ^2P_{1/2}$ & H$_2$O v=0 $3_{(2,2)}-3_{(1,3)}$ &  \\
\hline\hline
RA, DEC & (J2000) &  23:10:38.89, 18:55:19.8 & 23:10:38.90, 18:55:19.8 & 23:10:38.90, 18:55:19.8 \\
$F_{obs}$ & [GHz] & 271.382 & 274.074 & ...\\
z$_{line}$ & & 6.0031 $\pm$ 0.0001 & 6.0031 $\pm$ 0.0006 & ...\\
FWHM & [$\rm km\ s^{-1}$] & 422 $\pm$ 15$^a$ & 340 $\pm$ 88$^b$ & ...\\
$Sdv$ & [Jy $\rm km\ s^{-1}$] & 5.2 $\pm$ 0.01 & 0.36 $\pm$ 0.01 & ...\\
L$_{\rm line}$ & [$10^9\ L_{\odot}$] & 5.1 $\pm$ 0.1 & 0.36$\pm$ 0.01 & ...\\
S$_{\rm cont}$ & [mJy] & ... & ... & 6.43 $\pm$ 0.16 \\
size & [arcsec$^2$] & 0.449 $\times$ 0.325$^a$ &  0.332 $\times$ 0.192$^b$ & 0.225 $\times$ 0.190$^b$ \\
size & [kpc$^2$] & 2.6 $\times$ 1.9 & 1.9 $\times$ 1.1 & 1.3 $\times$ 1.1 \\
\hline                                   
\end{tabular}
 \flushleft 
 \footnotesize{{\bf Notes.} Line fluxes are derived by integrating over the line profiles extracted from the region included within $>2\sigma$ in the velocity integrated map. Sizes are estimated with a 2D Gaussian fit in CASA. $^a$ FWHM size with Briggs cleaning with robust=0.5. $^b$ FWHM size with natural cleaning.}
\end{table*}

\begin{table*}
\caption{Continuum}
\centering       
\label{table-sed}                        
\begin{tabular}{c c c c c c c c}        
\hline\hline  
 Frequency  & Synth. beam & r.m.s. & Flux density & Size & Project ID & Telescope & References \\ 
 (GHz) & [arcsec$^2$] & [mJy/beam] & [mJy] & [arcsec$^2$] & \\
 \hline
  91.500 & 0.71 $\times$ 0.43 & 0.0053 & 0.29 $\pm$ 0.01 & 0.261 $\times$ 0.171 & 2015.1.00584.S & ALMA & TP, [1] \\
  136.627 & 0.74 $\times$ 0.71 & 0.015 & 1.29 $\pm$ 0.03 & 0.345 $\times$ 0.212 & 2015.1.01265.S & ALMA & TP, [2], [3], [4] \\
  140.995 & 0.79 $\times$ 0.65 & 0.015 & 1.40 $\pm$ 0.02 & 0.263 $\times$ 0.212 & 2015.1.01265.S & ALMA & TP, [2], [3], [4] \\
 153.070 & 0.21 $\times$ 0.17 & 0.0091 & 1.63 $\pm$ 0.06 & 0.214 $\times$ 0.189 & 2018.1.00597.S & ALMA & This paper \\
 263.315 & 0.14 $\times$ 0.11 & 0.016 & 7.73 $\pm$ 0.31 & 0.190 $\times$ 0.180 & 2018.1.00597.S & ALMA & This paper \\
 265.369 & 1.62 $\times$ 1.3 & 0.056 & 8.81 $\pm$ 0.13 & 0.456 $\times$ 0.422 & 2019.1.01721.S & ALMA & This paper \\
 284.988 & 0.51 $\times$ 0.39 & 0.073 & 11.05 $\pm$ 0.16 & 0.233 $\times$ 0.220 & 2013.1.00462.S & ALMA & TP, [3] \\
289.180 & 0.58 $\times$ 0.47 & 0.025 & 11.77 $\pm$ 0.12 & 0.330 $\times$ 0.246 & 2015.1.01265.S & ALMA & TP, [2], [3], [4] \\
 344.185 & 0.53 $\times$ 0.43 & 0.051 & 14.63 $\pm$ 0.34 & 0.289 $\times$ 0.229 & 2015.1.01265.S & ALMA & TP, [2], [3], [4] \\
 490.787 & 0.7 $\times$ 0.6 & 0.10 & 25.31 $\pm$ 0.19 & 0.318 $\times$ 0.229 & 2017.1.01195.S & ALMA & This paper, [5] \\
 599.584 & ... & ... & <29.4 & ... & ... & \textit{Herschel}/SPIRE & [2] \\
 856.549 & ... & ... & 22.0 $\pm$ 6.9 & ... & ... & \textit{Herschel}/SPIRE & [2] \\
 1199.169 & ... & ... & 19.9 $\pm$ 6.0 & ... & ... & \textit{Herschel}/SPIRE & [2] \\
 1873.703 & ... & ... & 13.2 $\pm$ 2.8 & ... & ... & \textit{Herschel}/PACS & [2] \\
 2997.924 & ... & ... & 6.5 $\pm$ 1.2 & ... & ... & \textit{Herschel}/PACS & [2] \\

\hline
\end{tabular}
 \flushleft 
 \footnotesize{{\bf Notes.} All ALMA observations listed in this table are archival, and we have analysed them, even if the data-set was already been studied (see Ref. column). References: This paper (TP); [1] \citet{feruglio2018}; [2] \citet{shao2019}; [3] \citet{carniani+19}; [4] \citet{li2020}; [5] \citet{hashimoto2019}.}
\end{table*}

\begin{figure*}
        \centering
        \includegraphics[width=1\linewidth]{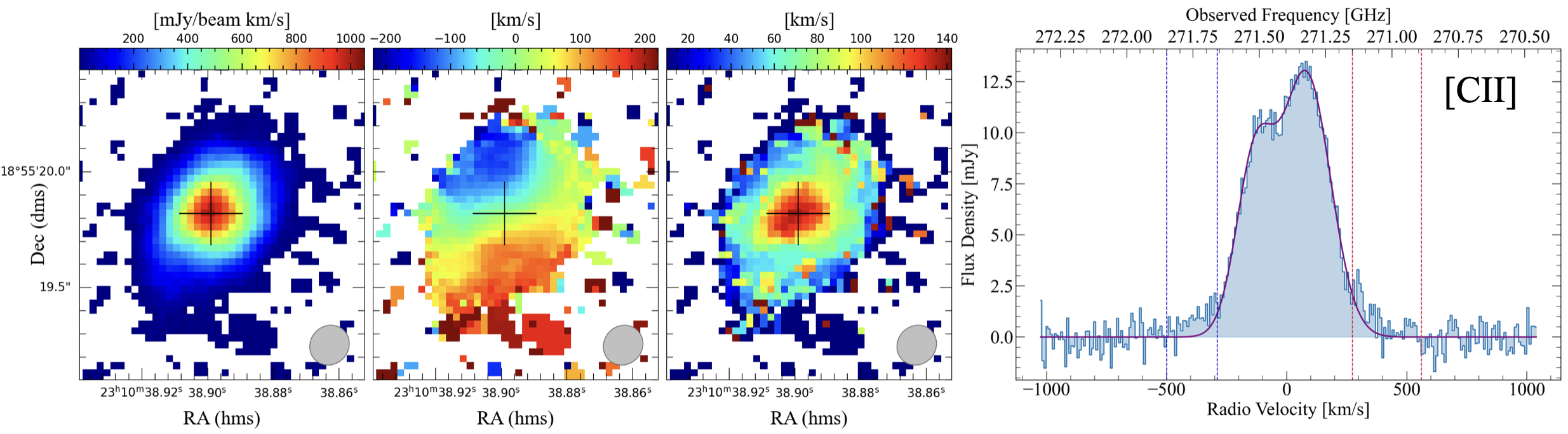}
                \includegraphics[width=1\linewidth]{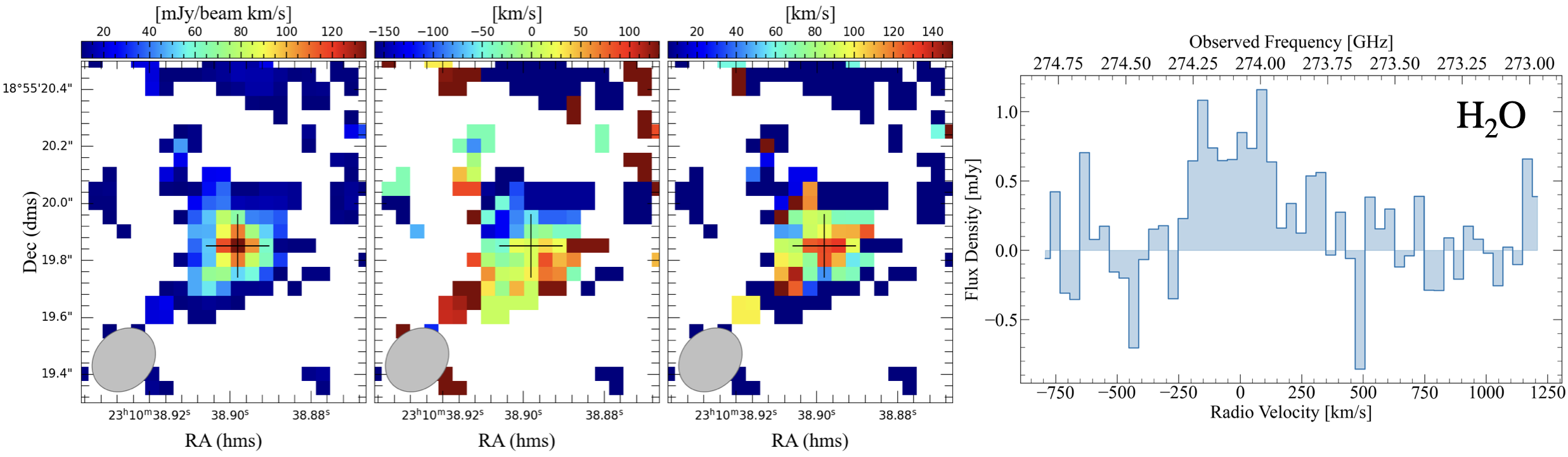}
        \caption{Moment maps of the [CII] emission line (top panels) and of the emission  line detected at 274.084 GHz (bottom panels), identified as the H$_2$O v=0 3(2,2)-3(1,3). From left to right: integrated flux, mean velocity map, and velocity dispersion map, continuum-subtracted spectra of [CII] (top right) and H$_2$O (bottom right). The clean beam is plotted in the lower right or left corner of the moment maps. The cross indicates the peak position of the integrated flux for each line.
        The spectra have been extracted from the region included within $\geq 2\sigma$ in the velocity integrated map. In the [CII] spectrum, the vertical blue and red lines highlight the spectral regions in which the flux is higher than in the Gaussian fit (solid purple line). The H$_2$O spectrum has been rebinned to 40 km s$^{-1}$.}
        \label{momj2310}
\end{figure*}

\begin{figure*}
        \centering
        \includegraphics[width=1\linewidth]{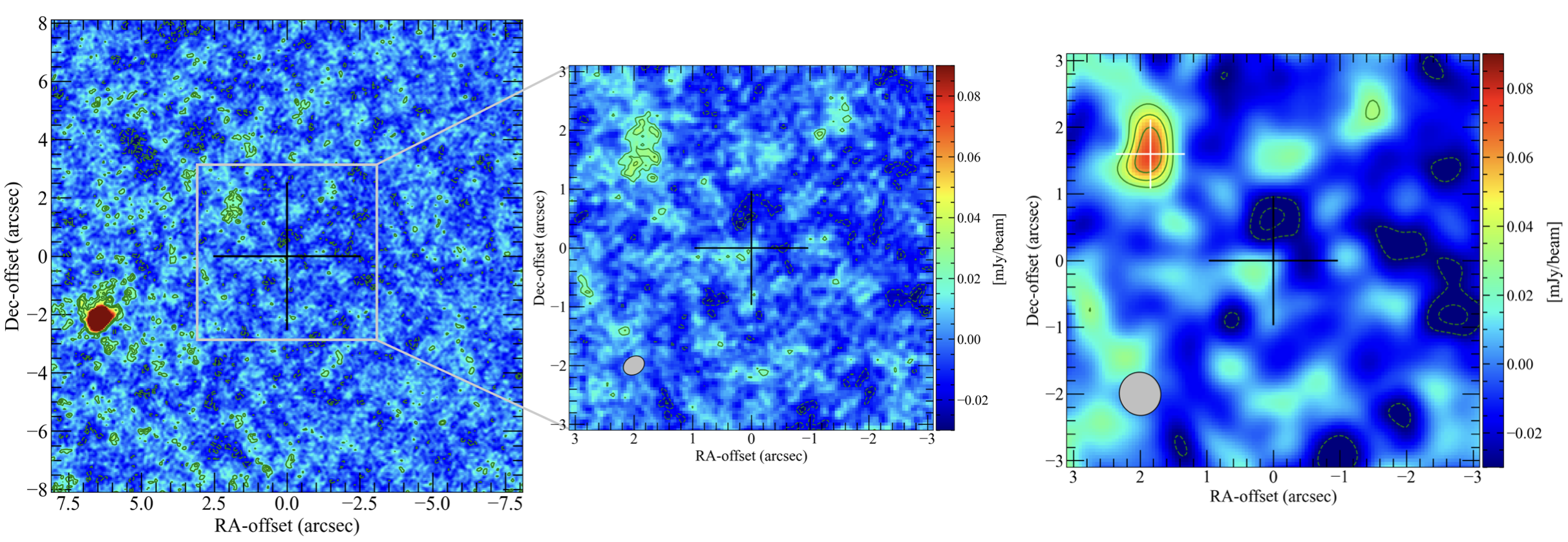}
        \caption{ Dust continuum maps of the QSO field with different resolutions. Left panel: dust continuum map of the QSO field centred on the phase-tracking centre of the observation. Levels are $-3,-2,2,3,\text{and }4\sigma$, $\sigma = 8.8\ \mu$Jy/beam. The cross indicates the phase centre.  In this map the QSO is located at offset  [6,-2] arcsec.  Central panel: Zoomed view towards the phase centre. The clean beam is plotted in the lower left corner.  Right panel: 260 GHz dust continuum uv-tapered map of Serenity-18 (levels $-4,-3,-2,2,4,\text{and }5\sigma$, $\sigma = 15\ \mu$Jy/beam). The imaging has been performed with \textit{uvtaper}=[$0.5$ arcsec]. The clean beam is indicated in the lower left corner. The black cross indicates the position of the phase-tracking centre, which coincides with the expected position of the CO-emitter Serenity-18 \citep{dodorico2018}. The white cross indicates the continuum emitter detected in the beam (see text).}
        \label{cont_sere}
\end{figure*}

\section{Results}
\label{sec:res}

\subsection{QSO continuum emission}

Figure \ref{contj2310} shows the 260 GHz dust continuum map obtained through natural weighting. The continuum shows resolved emission with an approximate size of 1.5 arcsec across, corresponding to 8.7 kpc at the rest frame. 
Using a 2D Gaussian fit, we derived a flux density of $6.43\pm 0.16$ mJy, 30\% lower than the value reported by \citet{shao2019} from lower-resolution data, and a FWHM size of $0.22 \times 0.19$ arcsec$^2$ (Table \ref{table:misure}). We note, however, that the Gaussian 2D fitting procedure fails to fit the surface brightness distribution and shows strong residual flux. Integrating the flux in the map over the region with $>2\sigma$, we derived a flux density of $7.12\pm 0.2$ mJy. This flux density is still lower than that reported previously by \citet{shao2019}, meaning that the high-resolution data miss about $25\%$ of the flux. 
 Measurement of the flux density from dataset II, that is, at lower resolution, with a 2D Gaussian fit, indeed led to a value of $8.81\pm 0.13$ mJy, which is in agreement with the measurement reported by \citet{shao2019}. To be conservative, we therefore used the flux density from dataset II to study the dust continuum SED (see Sect. \ref{sec:dust}).
Table \ref{table-sed} reports the continuum data from a reanalysis of ALMA archival data and Herschel data from \citet{shao2019}. 

\subsection{[CII] and H$_2$O emission}
\label{sec:cII-h2o}

We used the continuum-subtracted data cube to study the [CII] line emission of the QSO. In order to study the [CII] kinematics, we adopted the Briggs cleaned data cube that enhances the angular resolution. Figure \ref{momj2310} (top panels) shows the moment-0, -1, and -2 maps of the [CII] emission and the spectrum, obtained by applying a 3$\sigma$ threshold to the Briggs clean cube. The [CII] distribution is spatially resolved with a size of $(0.449 \times 0.325) \pm (0.028 \times 0.021)\ \rm arcsec^2$ estimated from a 2D Gaussian fit on the velocity-integrated map (see Table \ref{table:misure}), and it shows a velocity gradient oriented north-east to south-west  with $\Delta v=400$ $\rm km\ s^{-1}$. The moment-2 map shows a range of the velocity dispersion between 20 and 140 $\rm km\ s^{-1}$, where the maximum value towards the nucleus is affected by beam smearing \citep{davies2011}. 

Figure \ref{momj2310} (top right panel) shows the continuum-subtracted [CII] line profile, which peaks at a frequency of 271.382 GHz, corresponding to $z=6.0031 \pm 0.0001$, consistent with previous determinations (e.g. \citealt{wang2013, shao2019}, see Table \ref{table:misure}).  The FWHM of the line is $422\pm15$ $\rm km\ s^{-1}$, derived from the fit with a single Gaussian.
Because the [CII] line shows two peaks, the value of the integrated flux would be $\sim 10\%$ overestimated using a single Gaussian. Therefore, we fitted the line profile with two Gaussian functions and derived an integrated flux of $[5.2\pm 0.01]$ Jy $\rm km\ s^{-1}$. This is $40\%$ lower than the flux reported by \citet{feruglio2018}, obtained using a dataset with a clean beam of about 0.9 arcsec. This suggests that these higher-resolution observations filter out part of the flux. The [CII] profile shows excess emission with respect to the best fit at the red and blue sides of the line, which would require additional Gaussian components. We discuss these high-velocity emissions in Sect. \ref{sec:ciikinematics}.

We detected an emission line at a sky frequency of 274.074 GHz towards the QSO position with a statistical significance of 10$\sigma$ and an integrated flux of $S_{\nu}dv=0.36\pm0.01$ mJy $\rm km\ s^{-1}$ (Table \ref{table:misure}). The line width is $340\pm88$ $\rm km\ s^{-1}$, consistent with the [CII] width. 
We identified this line as the transition of water vapour H$_2$O v=0 $3_{(2,2)}-3_{(1,3)}$ with F$_{\rm rest}=1919.359$ GHz, and derive a $z_{\rm H_2O}=6.0031 \pm 0.0006 $, consistent with the [CII] redshift of the QSO. 
Figure \ref{momj2310} (bottom panels) shows the moment maps of this emission line, obtained through natural weighting to maximise sensitivity, and the line profile.
The emission appears to be spatially resolved with size of $1.94\times 1.12$ kpc$^2$ (Table \ref{table:misure}). A velocity gradient is detected
along a position angle (PA) that appears to be consistent with that detected in the [CII] line. 

\begin{figure*}
        \centering
        \includegraphics[width=1\linewidth]{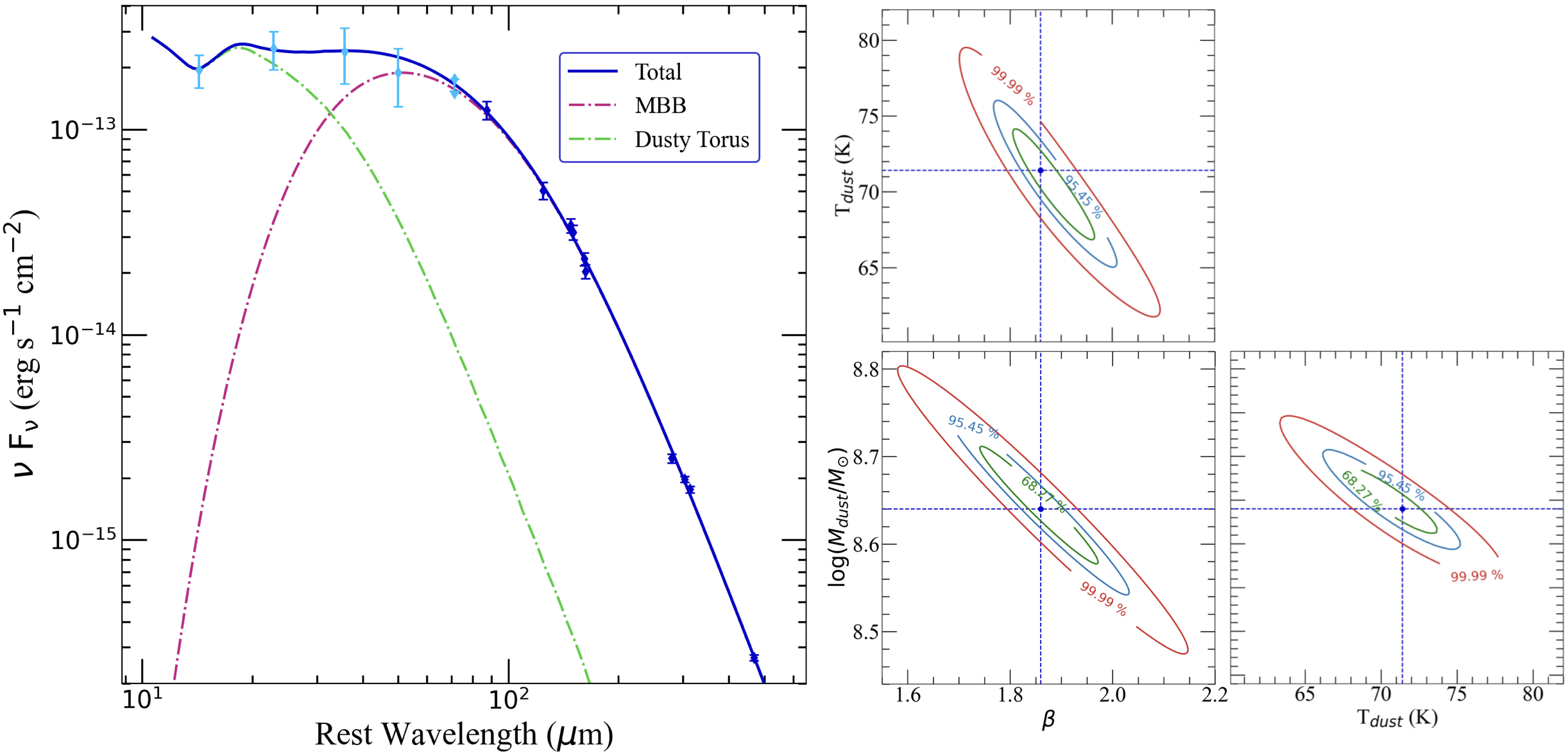}
        \caption{ Results of the SED fitting of J2310+1855. Left panel: SED of J2310+1855 using ALMA and Herschel data reported in Table \ref{table-sed}, fitted with an MBB function (for the dust emission) and the $a = 0.0002, \tau=3, \rm p=0, q=1.5, oa=80, R=30,\text{ and } i=20$ deg dusty torus model from the SED library SKIRTOR \citep{stalevski+2016}. Dark blue diamonds are computed based on ALMA observations and light blue diamonds based on Herschel observations, as listed in Table \ref{table-sed}. Right panel: Confidence ellipses for the MBB free parameters $T_{\rm dust}, M_{\rm dust},\text{ and } \beta$ computed at $68.27\%, 95.45\%,$ and $99.99\%$ confidence levels. The dotted blue lines indicate the best-fit parameters.}
        \label{sed}
\end{figure*}

We scanned the data cubes in each of the four spectral windows with the aim of searching for any additional line emitters within the ALMA beam. To do this,  we used a detection threshold of $S/N>3$ per spectral channel, and we required $S/N>3$ at the peak position of each detected structure over at least 20 adjacent channels (i.e. $\Delta v \gtrsim 150\rm ~km\ s^{-1}$). We did not detect any sources other than those described above.

One continuum emitter was detected at the 4$\sigma$ significance level at position (RA, DEC) = (23:10:38.57, +18:55:23.55) (Fig. \ref{cont_sere}) after we had downgraded the resolution of our observation applying \texttt{uvtaper}=[$0.5\ \rm arcsec$] within \texttt{tclean}. We obtained a clean beam of $0.65 \times 0.62$ arcsec$^2$ for the continuum map. The distribution of this structure is spatially resolved with a size of $(0.918 \times 0.125) \pm (0.194 \times 0.111)$ arcsec$^2$ and a flux of $0.137 \pm 0.015$ mJy, estimated from a 2D Gaussian fit. Another scan of the data cube yielded no additional continuum or line emitters in the field above the detection threshold of $S/N>1$ per channel. 

\section{Analysis}
\label{sec:disc}

\subsection{Dust properties}
\label{sec:dust}

In Fig. \ref{sed} we show the mm to FIR SED of J2310+1855 starting from a rest frame wavelength of 10 $\mu$m derived from the measurements in Table \ref{table-sed}. 
In this wavelength range, two main components contribute to the QSO emission:  the large-scale dust in the ISM, and the dusty torus. Following \cite{carniani+19}, we modelled the SED of the dust emission with a modified black-body (MBB) function given by
 \begin{equation}\label{eqsed}
     S_{\nu_{\rm obs}}^{\rm obs} = S_{\nu/(1+z)}^{\rm obs} = \dfrac{\Omega}{(1+z)^3}[B_{\nu}(T_{\rm dust}(z))-B_{\nu}(T_{\rm CMB}(z))](1-e^{-\tau_{\nu}}),
 \end{equation}

\noindent where $\Omega = (1+z)^4A_{\rm gal}D_{\rm L}^{-2}$ is the solid angle with $A_{\rm gal}$ , and $D_{\rm L}$ is the surface area and luminosity distance of the galaxy, respectively. The dust optical depth is
\begin{equation}
    \tau_{\nu}=\dfrac{M_{\rm dust}}{A_{\rm galaxy}}k_0\biggl(\dfrac{\nu}{250\ \rm GHz}\biggr)^{\beta},
\end{equation}
\noindent with $\beta$ the emissivity index and $k_0 = 0.45\  \rm cm^{2}\ g^{-1}$ the mass absorption coefficient \citep{beelen+2006}. The solid angle is estimated using the continuum emission mean size of the ALMA observations in Table \ref{table-sed}. The effect of the CMB on the dust temperature is given by
\begin{equation}
    T_{\rm dust}(z)=((T_{\rm dust})^{4+\beta}+T_0^{4+\beta}[(1+z)^{4+\beta}-1])^{\frac{1}{4+\beta}},
\end{equation}
\noindent with $T_0 = 2.73$ K.
We also considered the contribution of the CMB emission given by $B_{\nu}(T_{\rm CMB}(z)=T_0(1+z))$ \citep{dacunha2013}. Applying Eq. \ref{eqsed}, we performed a fit of ALMA data using a non-linear least-squares fit. We set $T_{\rm dust}, \log(M_{\rm dust}/M_{\odot}), \beta$ as free parameters, varying in the intervals $20\ {\rm K} \lesssim T_{\rm dust}\lesssim 300\ {\rm K}$, $6\lesssim \log(M_{\rm dust}/M_{\odot})\lesssim 10$, and $1\lesssim \beta \lesssim 2$ because these are reasonable ranges for high-z QSOs. The best-fit model has $\beta=1.86\pm 0.11$, a dust temperature $T_{\rm dust}=72\pm4$ K, and a dust mass of $M_{\rm dust}=(4.4\pm0.5)\times 10^8$ M$_\odot$. 

\begin{table}
\caption{Results of the SED fitting with MBB and dusty torus models}
\centering       
\label{table-sed-res}                        
\begin{tabular}{l c c}        
\hline\hline  
 \multicolumn{3}{c}{Dust emission} \\
\hline\hline
$\log(M_{\rm dust}/M_{\odot})$ & & 8.64 $\pm$ 0.07 \\
$M_{\rm dust} $ & $[10^8\ M_{\odot}]$ & 4.4 $\pm$ 0.7 \\
$T_{\rm dust}$ & [K]  & 71 $\pm$ 4\\
$\beta$ & & 1.86 $\pm$ 0.12 \\[0.1cm] 
$L_{\rm TIR, MBB}$  & [$10^{13}\ \rm L_{\odot}$] & 2.48$^{+0.62}_{-0.52}$\\ [0.1cm] 
$L_{\rm TIR, MBB+Torus}$ & [$10^{13}\ \rm L_{\odot}$] & 8.44$^{+0.62}_{-0.52}$ \\ [0.1cm] 
GDR$^a$ & & $101\pm20$ \\

\hline                                   
\end{tabular}
 \flushleft 
 \footnotesize{{\bf Notes}. SED fits are performed using two components, MBB and dusty torus (see Sect. \ref{sec:dust}). The table reports the individual contribution of the MBB component and the global contribution of MBB+torus for TIR (8-1000 $\mu$m).
 $^a$Gas-to-dust mass ratio derived from $M_{\rm dust}$ (this work) and molecular mass $M(\rm H_2)=(4.4\pm 0.2)\times 10^{10}\ \rm  M_\odot$ derived from CO(2-1) and (6-5) \citep{li2020,feruglio2018}.} 

\end{table}

The flux excess with respect to the MBB that is probed by Herschel photometric points requires a warmer dust component, which we modelled with a dusty torus component. 
We used SKIRTOR, a library of SED templates to model the AGN dusty torus, calculated with SKIRT, a custom radiative transfer code based on Monte Carlo techniques \citep{stalevski+2016}. This library is made of 19200 templates with different values of the optical depth $\tau$ at $9.7\  \mu$m,  of the power-law exponent that sets the radial gradient of dust density ($p$), of the index for dust density gradient with polar angle ($q$), eight different half-opening angle OAs between the equatorial plane and the edge of the torus, ten inclinations $i$, from face-on ($0$ deg, for typical unobscured type I AGN) to edge-on ($90$ deg, obscured type II AGN) view, and three values for the ratio $R$ of the outer to inner radius of the torus.

Thus, we used Eq. \ref{eqsed} of the MBB and the torus templates to fit the flux continuum densities measured with ALMA and Herschel. We set the $T_{\rm dust}$, $\log(M_{\rm dust}/M_{\odot})$, $\beta$, and $a$ (the normalisation of the torus template) as free parameters and explored the parameter space using a non-linear least-squares fit. We forced $T_{\rm dust}$, $\log(M_{\rm dust}/M_{\odot})$, $\beta$ as before, and $10^{-5}\lesssim a \lesssim 1$. 

Figure \ref{sed} shows the results of the SED modelling with a combination of a dusty torus and an MBB, and the $\chi^2$ confidence contours for $T_{\rm dust}$, $\log(M_{\rm dust}/M_{\odot})$, $\beta$.  Table \ref{table-sed-res} reports the best-fitting results. The best-fitting parameters for the MBB are $\beta=1.86\pm 0.12$, a dust temperature $T_{\rm dust}=71\pm4$ K, and a dust mass of $M_{\rm dust}=(4.4\pm0.7)\times 10^8$ M$_\odot$. These values are consistent with those obtained from the fit with the MBB alone, implying that the warmer dusty torus component has little impact on the cool dust component from the host galaxy.
Our best-fit $T_{\rm dust}$ is a factor of $\sim2 $ higher than that derived by \citet{shao2019} ($T=39$ K), and $M_{\rm dust}$ is a factor of $\sim4$ smaller than their estimate. This also implies a higher 
gas to dust mass ratio compared to their estimates. We find ${\rm GDR}=101\pm20$ based on our $M_{\rm dust}$ estimate and the molecular mass  $M(\rm H_2)=(4.4\pm 0.2)\times 10^{10}\ \rm  M_\odot$, measured from CO(2-1) and CO(6-5) by \citet{li2020} and \citet{feruglio2018}, using the commonly adopted conversion factor for QSO host galaxies $\alpha_{\rm CO}=0.8\rm ~M_{\odot} ~(K ~km ~s^{-1} ~pc^2)^{-1}$ \citep{downes1998, carilli2013}.

We note that the photometric data at $\lambda_{\rm rest}<15\mu$m can be equivalently well fitted by a broad range of SKIRTOR templates (that differ by $\Delta \chi^2 = 0.01$ at most), depending on the combination of the torus parameters. In particular the torus inclination, $i$, is a great source of degeneracy: when the other torus parameters are fixed to the best-fitting values, the variation in $i$ yields a set of equivalent templates, with $\Delta\chi^2 = 0.01$. The inclination of the torus can be properly determined by observations in the wavelength regime $\lambda_{\rm rest}<10 \mu$m, which is not covered by our dataset. Although the contribution of the dusty torus, clearly seen in the flux excess at $\lambda_{\rm rest}<15\mu$m, should be considered in the fitting procedure to properly characterize the physics of the QSO and its host galaxy, we were not able to uniquely determine the structural and physical properties of the dusty torus with this dataset. Nonetheless, we were able to use the best-fitting function MBB+template to compute the total infrared (TIR) luminosity from $8$ to $1000\ \mu$m rest-frame, retrieving a value of $L_{\rm TIR} = 8.44^{+0.62}_{-0.52}\times 10^{13}\ \rm L_{\odot}$ (see Table \ref{table-sed-res}). 

We also estimated the TIR luminosity for the best-fit MBB model by integrating from $8$ to $1000\ \mu$m rest-frame, and we obtained $L_{\rm TIR} = 2.48^{+0.62}_{-0.52}\times 10^{13}\ \rm L_{\odot}$ \citep{duras2017}. Several observations and radiative transfer simulations suggested that the radiative output of luminous QSOs substantially contributes to dust heating on kpc scale \citep{duras2017, dimascia2021, bischetti2021, walter2022}. In particular, \citet{duras2017} showed that about $50\%$ of the total IR luminosity in AGN with $L_{\rm bol}>10^{47}\rm \ erg ~s^{-1}$ is due to dust heated by QSOs. Applying this correction and adopting a Chabrier initial mass function (IMF; \citealt{chabrier2003}) would imply a ${\rm SFR} = 1240^{+310}_{-260}\ \rm M_{\odot}yr^{-1}$ and a $\Sigma_{\rm SFR} = 521\ \rm M_{\odot}yr^{-1}kpc^{-2}$,  within the dust half-light radius ($0.87$ kpc, see Sect. \ref{sec:ciikinematics} for the detailed derivation of the half-light radius).

\begin{figure*}
        \centering
        \includegraphics[width=0.95\linewidth]{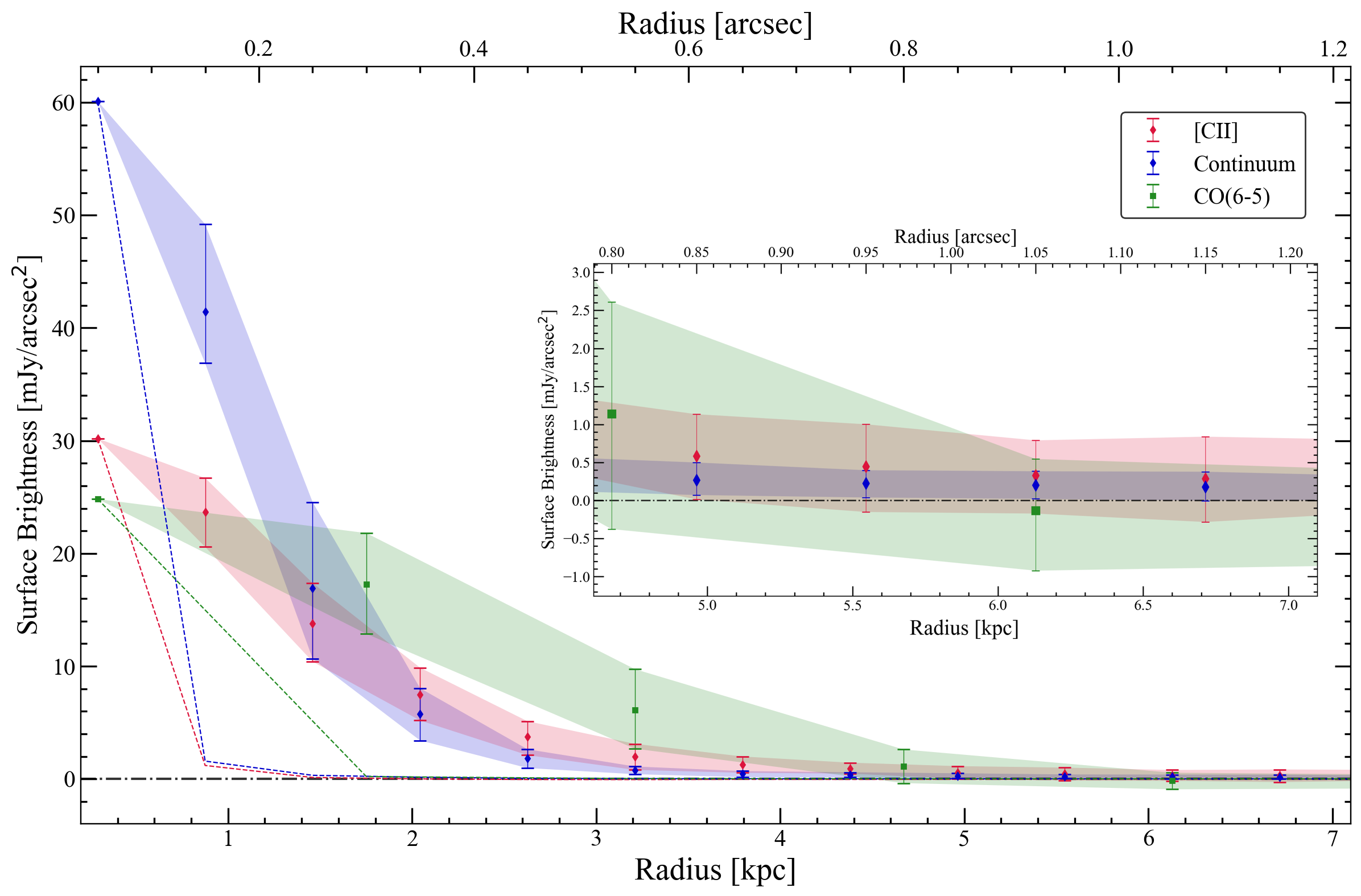}
        \caption{Natural weighting surface brightness for [CII] (red), CO(6-5) (green), and continuum (blue) of J2310, normalised to the respective r.m.s. level,  as a function of the radius from the peak position of the source. The lower and upper error bars are the 16th and 84th percentiles, respectively. The shadowed regions connect the uncertainties given by the percentiles. The coloured dashed lines show the synthetic beams for each observation. The dashed black lines mark the 0 level of surface brightness. The inset shows a zoomed view at large radii. The CO(6-5) data are taken from \citet{feruglio2018} and have a clean beam of $0.6\times 0.4$ arcsec$^2$.}
        \label{sbright-prof}
\end{figure*}

\subsection{[CII] distribution and kinematics}
\label{sec:ciikinematics}

\begin{figure*}
        \centering
        \includegraphics[width=0.9\linewidth]{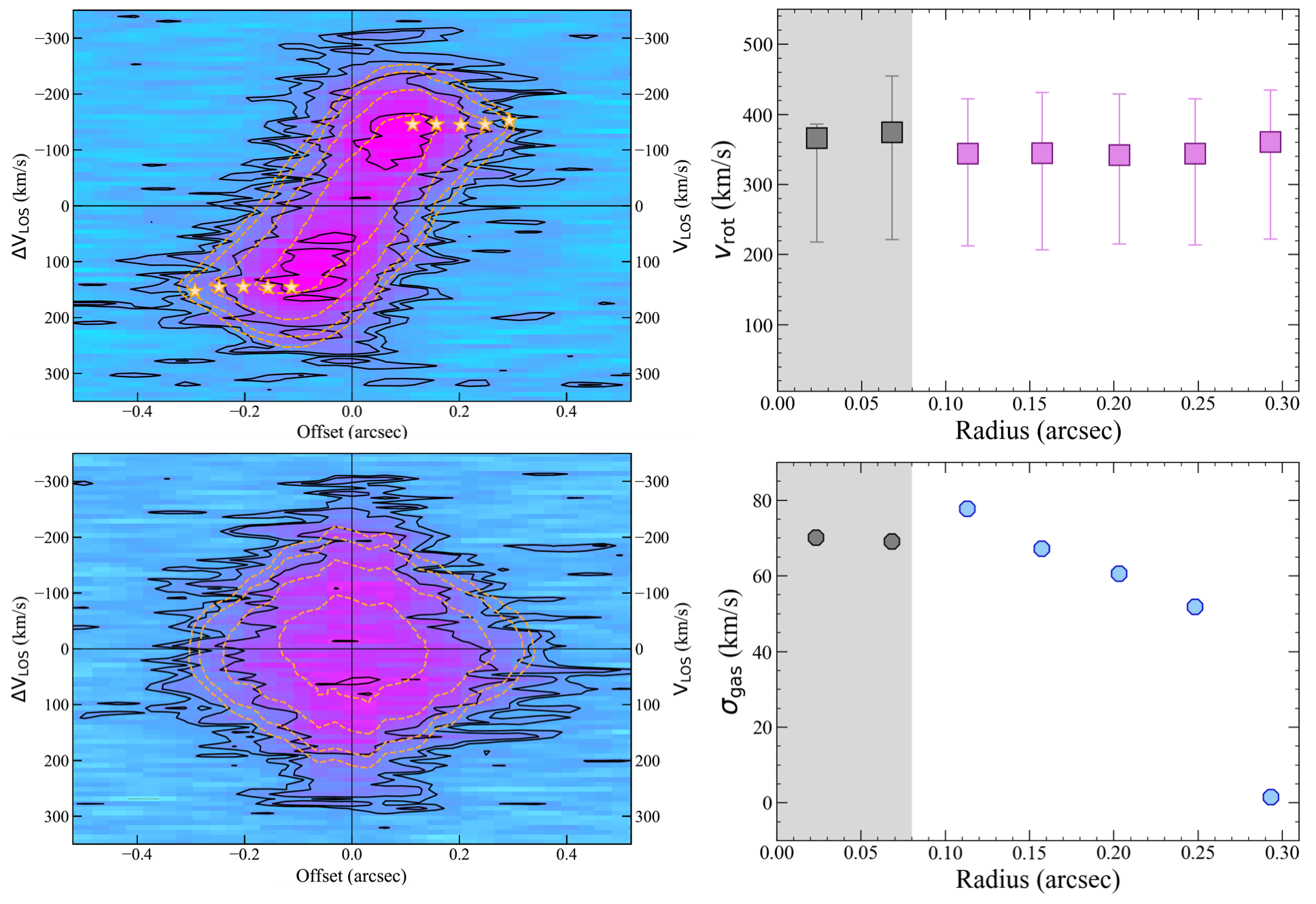}
        \caption{ Results of the $^{\rm 3D}$Barolo dynamical modelling of the [CII] emission line. Top and bottom left: PV diagrams of the [CII] emission line along the line of nodes (PA=$200$ deg) and orthogonal to it (PA=$290$ deg), performed with $^{\rm 3D}$Barolo. Contours are at $2,3,6,\text{and }12 \sigma$, with $\sigma = 0.22$ mJy, for the data (solid black lines) and the best-fit model (dashed orange lines). Sampling is performed with five radii, spaced by $0.045$ arcsec. Yellow stars show the disk model. Top and bottom right: Rotational velocity, $v_{\rm rot}$, and gas  velocity dispersion, $\sigma_{\rm gas}$, as a function of radius for the best-fit disk model. To compute the top and bottom error bars for the rotational velocity, we performed the modelling by varying the disk inclination $i= 20 \text{ and }45$ deg, respectively.  The grey shaded area marks the central beam ($r_{\rm beam} \sim 0.08$ arcsec), and grey points are those computed including the central beam.} 
        
        \label{pvdiagram}
\end{figure*}

We measured a total line luminosity of $L_{\rm [CII]}=5.1\pm0.1 \times 10^9\ \rm L_{\odot}$ and derived an $L_{\rm [CII]}/L_{\rm TIR, MBB+Torus}=6\times 10^{-5}$. This value is at the lower end of the distribution found for QSOs at this redshift \citep{walter2022, decarli2018,shao2019}. 
The neutral gas mass in the disk, based on [CII] emission, can be derived with the relation from \citet{Hailey2010} (see also \citealt{bischetti2019b}),

\begin{equation}
    \label{eq:cii_mass}
    \dfrac{M_{\rm HI}}{M_{\odot}} = 0.77\biggl(\dfrac{0.7L_{\rm [CII]}}{L_{\odot}}\biggr)\biggl(\dfrac{1.4\times 10^{-4}}{X_{C^+}}\biggr)\times\dfrac{1+2e^{-91{\rm K}/T}+n_{\rm crit}/n}{2e^{-91{\rm K}/T}}
,\end{equation}

\noindent where $X_{C^+}$ is the [CII] fraction per hydrogen atom, T is the gas temperature, $n$ is the gas density, and $n_{\rm crit}\sim 3 \times 10^3$ cm$^{-3}$  is the [CII]$\lambda$158$\mu$m critical density  for collisions with neutral hydrogen that frequently occur in photo-dissociation regions (PDRs; \citealt{wolfire2022,hollenbach1999}). 
 We estimated the lower limit for the molecular mass in the regime $n\gg n_{\rm crit}$ \citep{maiolino2005, aalto2012, aalto2015}, and we considered a $X_{C^+}\sim10^{-4}$ and a gas temperature of $200$ K, both typical of PDRs \citep{maiolino2005, Hailey2010, cicone2015, bischetti2019}.
We found $M_{\rm HI}=6.6\times 10^9~\rm M_{\odot}$, significantly lower than the molecular mass obtained through molecular tracers \citep{feruglio2018, shao2019}.

In Fig. \ref{sbright-prof} we show the observed surface brightness profiles of the [CII], dust continuum emission, and molecular gas traced by CO(6-5) (the latter taken from \citealt{feruglio2018}). To compute the profiles, we used natural weighted maps and defined concentric annular regions centred at the QSO continuum position peak (RA, DEC = 23:10:38.90, 18:55:19.8). The first point for the surface brightness was taken at the peak pixel, and the next annuli are $0.1$ arcsec thick for the [CII] and dust continuum, and $0.25$ arcsec thick for the CO(6-5). To compute the surface brightness at each annulus, we averaged the flux within each annulus and divided it by the area of each annulus. The error bars associated with the surface brightness at each annulus (plotted in Fig. \ref{sbright-prof}) mark the 16th and 84th percentiles inside each annulus, which corresponds to $\sim 68\%$ of the surface brightness distribution centred on the mean value (i.e. $\pm 1\sigma$ for a Gaussian distribution). To determine the rms level, we associated a Poisson error with each region that we obtained as follows. We computed the  rms per beam over a $10\times20$ arcsec$^2$ background, target-free area, and divided it by the square root of the number of beams in each annulus.  To compare the [CII], CO, and continuum profiles, which have different rms levels, we subtracted the rms from each respective profile, so that the three distributions approached zero at high radii. All profiles are more extended than their respective synthetic beam, shown as dashed lines in Fig. \ref{sbright-prof}. The [CII] and dust emission show consistent profiles for radii $r \gtrsim 1.5$ kpc.  In the inner region of the source, we found that the continuum is more peaked at the centre than [CII] (and CO). 
 The half-light radii are 0.87 kpc for the dust, 1.08 kpc for [CII], showing that the [CII] emission is more extended than the continuum emission (e.g. \citealt{li2022}), and 1.5 kpc for the CO emission. These results are also consistent with values that are commonly measured in QSOs at this redshift \citep{decarli2018,carniani+19,venemans2020}. A similar behaviour of reduced [CII] emissivity close to the QSO was observed in the $z\sim 7$ QSO J2348-3054 \citep{walter2022} and in the $z=6.6$ QSO J0305-3150 \citep{li2022}.  This is likely due to the contribution of the QSO to the dust heating, which reduces the $L_{\rm [CII]}/L_{\rm FIR}$ ratio in the nuclear region.
 We estimated the total gas surface density, including the contribution of HI from [CII] and $\rm H_2$ from CO within the half-light radius $r_{\rm [CII], HI}\sim 1.08$ kpc. We find $\Sigma_{\rm gas (H I + H_2)}=13809 \rm \ M_\odot ~pc^{-2}$. However, the gas traced by [CII] reaches a (azimuthally averaged) maximum size\footnote{The maximum size coincides with the radius at which the brightness profile reaches the zero level, within the error bars. It is shown more clearly in the zoomed panel of Fig. \ref{sbright-prof}.} of $r_{\rm [CII]}\sim5$ kpc, and the dust emission reaches an even larger radius of $r_{\rm dust}=6.7$ kpc, probably owing to the better sensitivity reached in the aggregated bandwidth. The molecular gas traced by CO(6-5) shows a smoother profile and reaches a size similar to the [CII], $r_{\rm CO}\sim 4.7$ kpc.

We modelled the [CII] line-of-sight (LOS) velocity distribution with an inclined-disk model, using the 3D-based analysis of rotating objects from line observations ($^{\rm 3D}$Barolo), a software package for the fitting of 3D tilted-ring models to emission line observations from high-resolution to very low resolution data cubes \citep{diteodoro2015}. $^{\rm 3D}$Barolo builds a number of models in the form of artificial 3D observations and compares them with the input cube, finding the set of geometrical and kinematical parameters that better describes the data, correcting for beam-smearing effects. Fixed parameters of the fit are the centre of the disk, set to the continuum peak (Table \ref{table:misure}), and the position angle PA$= 200$ deg. The rotational velocity $v_{\rm rot}$ and velocity dispersion $\sigma_{\rm gas}$ are free parameters with initial-guess values of $50$ $\rm km\ s^{-1}$ and $70$ $\rm km\ s^{-1}$, respectively. Since even in the case of high-resolution data the inclination, $i$, is the strongest source of uncertainty in determining the rotation curve and the dynamical mass, we performed a first run with $i$ set as a free parameter, retrieving a value of $i\sim 25$ deg for the best-fitting disk model. In the second run, we fixed the inclination to  $i=25$ deg, with all the other parameters as before. The sampling was initially performed with seven radii, spaced by $0.045$ arcsec, starting from a galactocentric radius of $0.023$ arcsec. This produced a rotation curve with an average $v_{\rm rot}\simeq 354$ $\rm km\ s^{-1}$ and a flat profile, while we would expect a decreasing trend towards the galactic centre. The excess velocity in the central part of the rotation curve could be due to a residual beam-smearing effect or to an additional kinematic component in the centre. In order to avoid the systematics induced by the inclusion of innermost region and to obtain a reliable modelling of the disk, we excluded the central beam (radius $\sim 0.08$ arcsec) from the fit, and performed a sampling with 5 radii, spaced by $0.045$ arcsec, starting from $0.113$ arcsec.  The top and bottom left panels of Fig. \ref{pvdiagram} show the position-velocity (PV) diagrams of the disk along the major and minor kinematic axes  with contours of the disk model as dashed orange lines and the modelled LOS velocities as orange stars. Exploring the $v_{\rm rot}-i$ parameter space with the task \texttt{SPACEVAR} of $^{\rm 3D}$Barolo, we found that models with $i$ in the range $[20,45]$ deg give similar results in modelling the PV diagram and the velocity dispersion profile. The $v_{\rm rot}$ is most affected by the variation of $i$ because $v_{\rm rot}=v_{\rm LOS}/\sin(i)$, and in this case, models with $i<20$ lead to unreasonably high values for the velocity rotation ($v_{\rm rot}\gtrsim 500-1000 \rm ~km~s^{-1}$).
In the top and bottom right panels of Fig. \ref{pvdiagram}, the velocity rotation and velocity dispersion curves are presented, where the error bars arise from the lower and upper limit inclination ($i=20,45$ deg).
 The intrinsic (i.e. beam-smearing corrected) rotational velocity $v_{\rm rot}$ 
 shows a rather flat rotation curve, and the velocity dispersion $\sigma_{\rm gas}$ is boosted at $r\sim 0.1$ arcsec, reaching $\sim 80\ {\rm km\ s^{-1}}$, while it decreases to $50\ {\rm km\ s^{-1}}$ at larger radii. We found $v_{\rm rot}\simeq 347\ {\rm km\ s^{-1}}$ and $\sigma_{\rm gas}\simeq 60\ {\rm km\ s^{-1}}$ on average within $r<1.5$ kpc (values corrected for beam smearing). We derive $v_{\rm rot}/\sigma_{\rm gas}\sim 6$, indicating a disk that is rotationally supported. 

 The total dynamical mass enclosed within a radius r=$1.7$ kpc ($\sim 0.3$ arcsec) is $M_{\rm dyn} = 5.2^{+2.3}_{-3.2}\times 10^{10}\ M_{\odot}$, consistent with that derived from CO(6-5) in approximately the same region \citep{feruglio2018} and from previous [CII] lower-resolution observations \citep{wang2013}. The uncertainties on $M_{\rm dyn}$ were obtained by propagating the error for $v_{\rm rot}$ at R=$1.7$ kpc.  Since the gas mass measured from CO is $M_{\rm H_2}=4.2\times 10^{10}\ \rm M_{\odot}$ (see Sect. \ref{sec:ciikinematics}), dynamical models with $i>30$ deg, which imply $M_{\rm dyn}\lesssim 4\times 10^{10}\ M_{\odot}$, can be ruled out. This restricts the
range of possible values for the inclination to [20,30] deg, supporting the choice of $i=25$ deg in the dynamical modelling of the disk.
 
\begin{figure}
    \centering
    \includegraphics[width=1\linewidth]{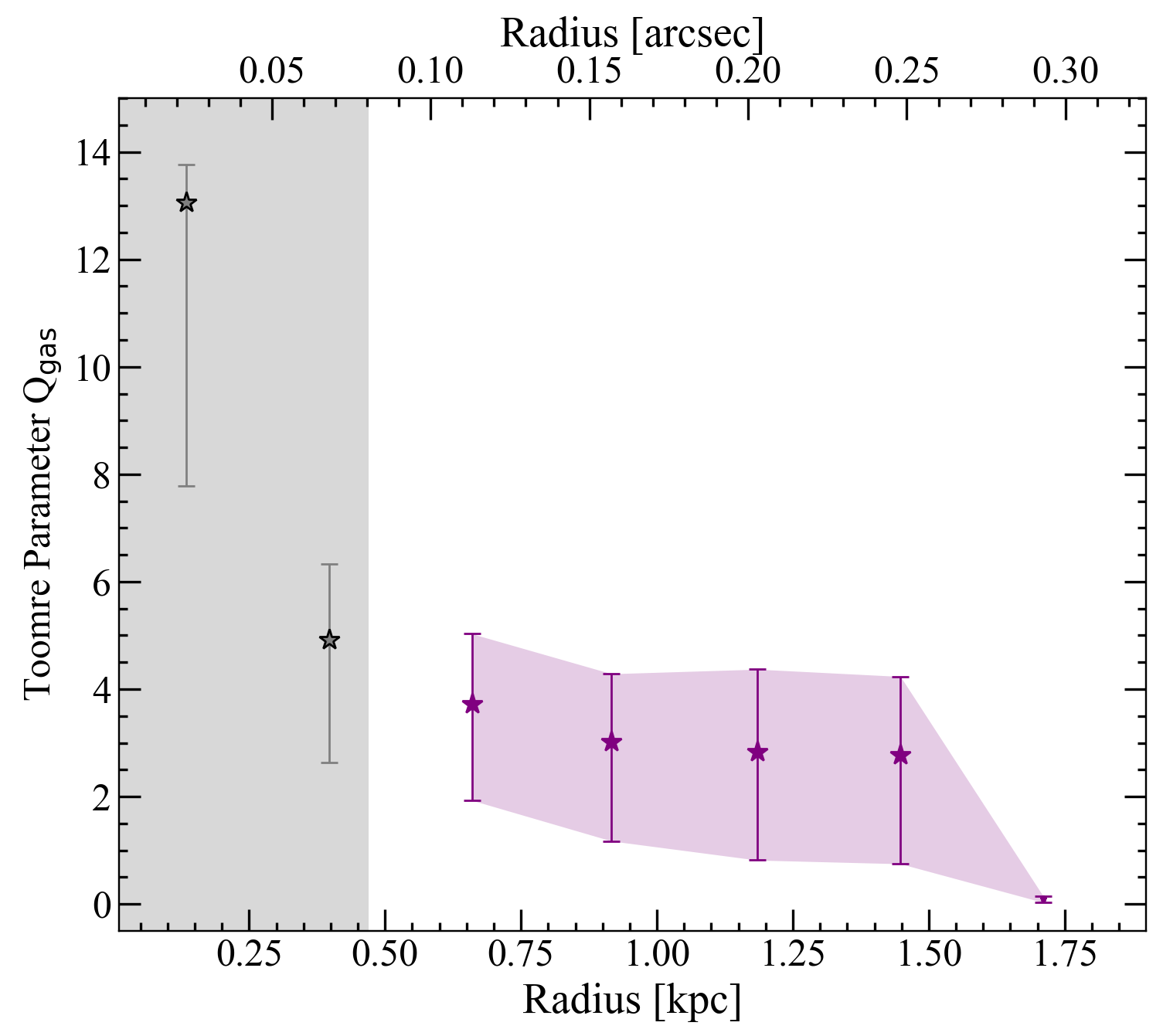}
    \caption{Gas Toomre parameter Q$_{\rm gas}$ as a function of the radius from the centre position of the [CII] emission, computed for the annuli used in the $^{\rm 3D}$Barolo [CII] modelling (see Sect. \ref{sec:ciikinematics} and Fig. \ref{pvdiagram}).  The grey shaded area marks the region of the central beam ($r_{\rm beam}\sim 0.08$ arcsec), and grey points are computed using the $v_{\rm rot}$ and $\sigma_{v}$ of the $^{\rm 3D}$Barolo model including the central beam.}
    \label{fig:toomre}
\end{figure}

The Toomre parameter, spatially averaged across the entire emission within $r<1.5$ kpc, is $Q_{\rm gas}=\sqrt{2}\sigma_v v_{\rm rot} / \pi G r \Sigma_{\rm gas} \approx 3$  for a flat rotation curve, where $\Sigma_{\rm gas}$ is the gas surface density derived from [CII] within the radius $r$. Considering also the Toomre parameter for the stellar component, the global $Q$ would be lower than this value \citep{aumer2010}.  
In Fig. \ref{fig:toomre} we show the radial profile of $Q_{\rm gas}$, where we computed the molecular gas surface mass density, taken in the same annuli as defined for the dynamical modelling (see Fig. \ref{pvdiagram}), and we propagated the uncertainties of $v_{\rm rot}$ and $\Sigma_{\rm gas}$.  The $v_{\rm rot}$ and $\sigma_{v}$ are those derived by the model of $^{\rm 3D}$Barolo at different radii (see top and bottom right panels of Fig. \ref{pvdiagram}). The grey shaded area, as before, marks the region of the central beam, and the grey points are computed using the $v_{\rm rot}$ and $\sigma_{v}$ of the $^{\rm 3D}$Barolo model including the central beam. $Q_{\rm gas}$ is in the range 3-13 in at $r< 0.4$ kpc, reflecting the flat rotation curve and high value of $v_{\rm rot}$ at the centre. For $r>0.8$ kpc, $Q_{\rm gas}\approx  1-5$ is close to the critical value commonly adopted for the gas component \citep{genzel2014, leroy2008}, indicating that the  disk is unstable against gravitational collapse, can fragment, and may eventually lead to star formation. 
$Q_{\rm gas}\sim 1$ was reported for QSO J234833.34–305410.0 at $z\sim 7$ \citep{walter2022}. 

\begin{figure*}
        \centering
        \includegraphics[width=0.9\linewidth]{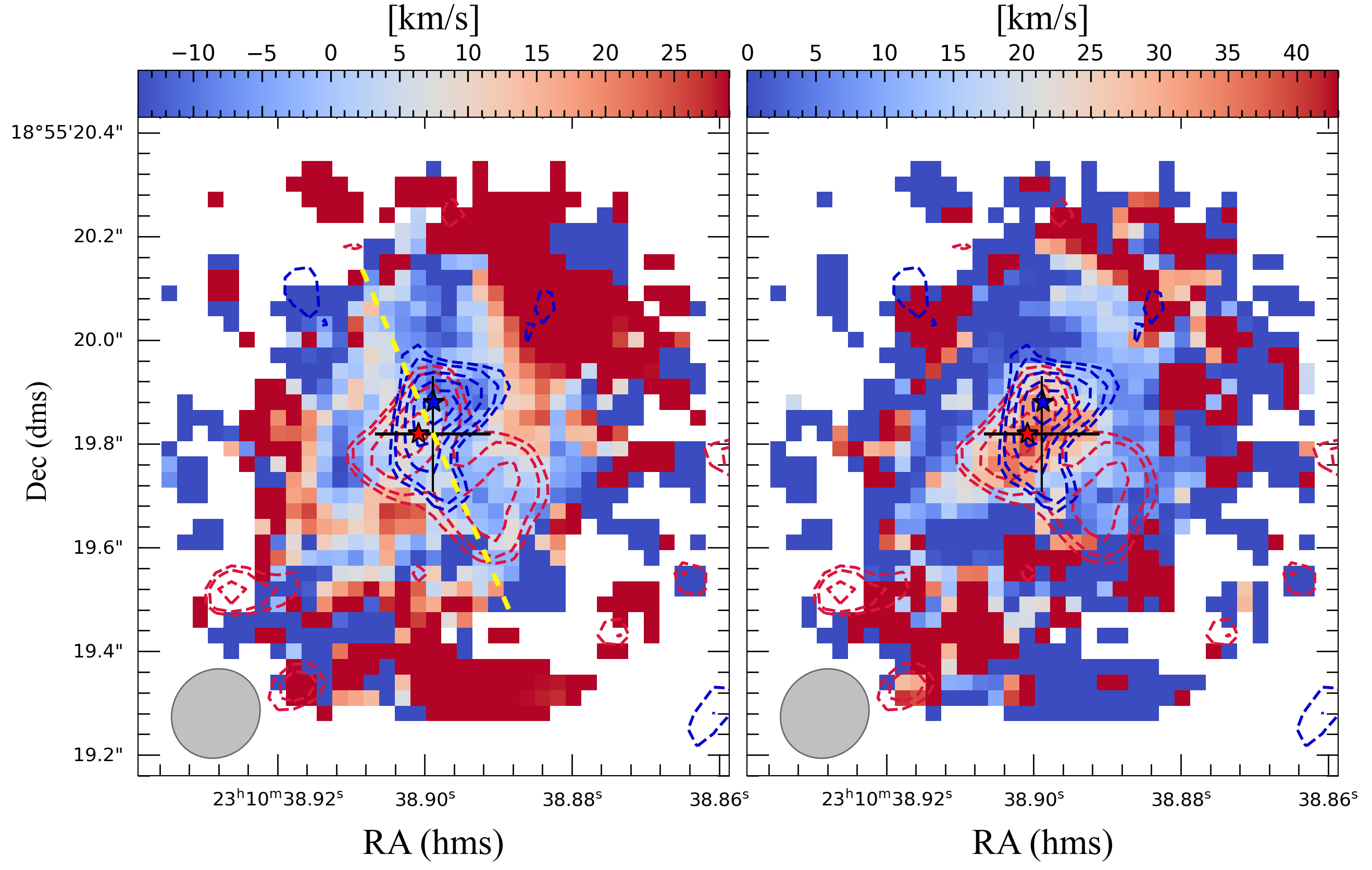}
        \caption{ $^{\rm 3D}$Barolo residuals of mean velocity map (left panel) and velocity dispersion map (right panel) of the [CII] emission line. The contours of red and blue wings are overplotted. Levels are $2.5,3,4,5,\text{and }6\sigma$, with $\sigma=39\ \mu$Jy for the red wing and $\sigma=43\ \mu$Jy for the blue wing. The red and blue wings have been selected from the [CII] spectrum as shown in Fig. \ref{momj2310}. The dashed yellow line is the kinematic major axis of the [CII] line, at PA = $200$ deg. The clean beam is indicated in the lower left corner. The excesses at the border of the source should not be considered because they are due to the poor subtraction of the noise.}
        \label{wings}
\end{figure*}

In Fig. \ref{wings}, the velocity-integrated contour maps of the blue and red wings, obtained by collapsing the spectral channels within the regions marked with blue and red lines in the [CII] spectrum of Fig. \ref{momj2310},  are shown overplotted on the $^{\rm 3D}$Barolo residuals of the [CII] velocity and velocity dispersion maps. The contour maps contain a contribution from the disk because the disk model is not subtracted from the data cube and from the corresponding maps, from which we extracted the contours. To evaluate the disk contribution to the wings, we used the [CII] spectrum and the double-Gaussian fit in Fig. \ref{momj2310}.  After subtracting the double-Gaussian components from the [CII] spectrum, we found that the integrated fluxes for the red and blue wings are $F_{\rm red} = 118$ mJy km s$^{-1}$ and $F_{\rm blue}=144$ mJy km s$^{-1}$ , respectively, evaluated by integrating the flux in the spectral regions marked by the vertical blue and red lines in the [CII] spectrum of Fig. \ref{momj2310}. Comparing these integrated fluxes with the total integrated flux in the same velocity range, we obtained that $F_{\rm red}/F_{\rm tot} \sim 65 \%$ and $F_{\rm blue}/F_{\rm tot} \sim 80 \%$. This implies that for the blue wing, the contribution of the disk is modest in the contours of Fig. \ref{wings}; for the red wing, the disk has a greater impact, but the contribution of the wing is still dominant. The dashed yellow line is the kinematic major axis, and the stars mark the peak positions of the emission of the blue and red wings.
These high-velocity emission regions are seen up to $3\sigma$ in the PV diagrams taken along the major and minor axes, within the offset $r \sim 0.1$ arcsec, and with an LOS velocity about $\pm 300\ {\rm km\ s^{-1}}$ on the blue and red sides of the line.  Along the minor kinematic axis, the disk dynamical model (orange contours) is not able to reproduce this excess emission, indicating that the latter is not produced by the beam-smearing effect, but is rather due to a different kinematic component. 

These high-velocity blue- and redshifted emissions located in the nuclear region do not follow the rotation curve of the main disk and may be due either to an unresolved circumnuclear disk tilted with respect to the main one, or to an outflow. 
 The high excess in LOS velocity between these components and the best-fit disk ($v_{\rm LOS, disk}\sim 100\ {\rm km\ s^{-1}}$) suggests that these blue- and redshifted emissions are due to an outflow and not to a tilted, unresolved nuclear disk.
We derived the flux density $S_{\nu}\Delta v$ of these two components by subtracting the double-Gaussian fit from the total [CII] spectrum, and integrating in the respective velocity ranges (see Fig. \ref{momj2310}). We computed the luminosity of the wings straightforwardly by applying Eq.1 from \citet{solomon2005} and their outflow mass using Eq. \ref{eq:cii_mass}. 
We obtained an outflow mass of  $M_{\rm out} = 1.9\times 10^8\ \rm M_{\odot}$ and $M_{\rm out} = 1.6\times 10^8\ \rm M_{\odot}$ for the blue and red wings, respectively. Assuming the scenario of time-averaged expelled shells or clumps \citep{rupke2005}, we computed the mass outflow rate for the blue and red wings,
\begin{equation}
    \Dot{M}_{\rm out} = \dfrac{v_{\rm out}\times M_{\rm out}}{R_{\rm out}}
,\end{equation}

\noindent where $v_{\rm out}$ is the projected outflow velocity defined as the velocity at which the integrated flux of each wing is 98\% of their total integrated flux with respect to the systemic velocity. We estimated  $v_{\rm out, blue}=-490$ $\rm km\ s^{-1}$ and $v_{\rm out, red}=535$ $\rm km\ s^{-1}$ from the line profile in Fig. \ref{momj2310}. We defined as outflow radius,  $R_{\rm out}$,  the projected separation between  the peaks of the red and blue wings and the [CII] peak position, that is, $R_{\rm out, blue} = 0.6$ kpc and $R_{\rm out, red} = 0.3$ kpc.  We then obtain an upper limit  $\Dot{M}_{\rm out} \lesssim 4500 \ \rm M_{\odot}\ yr^{-1}$ by adding the red and blue components, and adopting as outflow radius the maximum $R_{\rm out} (= 0.6$ kpc). However, depending on the LOS inclination, the intrinsic $R_{\rm out}$ may be larger than our estimate. Because these data do not allow resolving the kinematics of the outflow, an estimate of the lower limit for the mass outflow rate can be derived assuming an outflow size equal to the clean beam, $R_{\rm out}\sim 1$ kpc. This implies that a lower-limit outflow rate would be $\dot M_{\rm out}\gtrsim 1800\ \rm M_{\odot}\ yr^{-1} $.

We computed the kinetic power associated with the outflow as $\dot E_{\rm out} = \frac{1}{2} \dot M_{\rm out}\times v^2_{\rm out} =  (1.5-3.7)\times 10^{44} \rm erg ~s^{-1}$, and the wind momentum load
\begin{equation}
    \dfrac{\dot P_{\rm out}}{\dot P_{\rm AGN}} = \dfrac{\dot M_{\rm out}\times v_{\rm out}}{L_{\rm bol}/c}
,\end{equation}
\noindent where $\dot P_{\rm AGN}$ is the AGN radiation momentum rate. We adopted a bolometric luminosity of $L_{\rm bol}= 3.13\times10^{47}\rm erg ~s^{-1}$, derived from the rest-frame continuum at $3000$ \AA \citep{bischetti2022b} and the bolometric correction from \citet{runnoe2012}. 
This yields $\dot E_{\rm out}/L_{\rm bol} \sim 0.0005-0.001$ and $0.6 \lesssim \dot P_{\rm out}/\dot P_{\rm AGN} \lesssim 1.4$.

\subsection{$H_2O$ resolved emission}
\label{sec:h2oemission}

We identified the emission line detected at 274.074 GHz as the H$_2$O v=0 $3_{(2,2)}-3_{(1,3)}$ transition at rest frequency 1919.359 GHz. 
The emission line is detected with a $10 \sigma$ statistical significance, and the velocity gradient and PA are consistent with those derived for [CII]. 
Water vapour emission traces the molecular warm dense phase of the interstellar medium \citep{liu2017}, and is detected in only a few QSOs at z$>$6 \citep{Pensabene2021, Lehnert2020, Yang2019}. 
This is the first time that the emission is spatially resolved and consistent with a rotating water vapour disk. 
We estimated a first-order dynamical mass using H$_2$O. Adopting the same inclination as for the [CII] disk, i = 25 deg, we derived $M_{\rm dyn, H_2O}= 1.16\times 10^{5}(0.75 \times {\rm FWHM}_{\rm H_2O})^2 \times D  / \sin^2(i)=  6.4\times 10^{10}\rm ~M_\odot$, with ${\rm FWHM}_{\rm H_2O}= 340$ $\rm km\ s^{-1}$ and $D=1.5$ kpc (averaged source size in kiloparsecs, Table \ref{table:misure}). 
 
An alternative scenario for the identification of this line would be the [CII] line from the DLA J2310+1855 located at z=5.938646 (\citealt{dodorico2018}, see Sect. \ref{sec:env}). If the DLA had a line-emitting counterpart on the QSO sightline, its [CII] emission would happen at 273.906 GHz, which is also consistent with the observed line frequency. However, because the velocity gradient, the FWHM of the line, and the dynamical mass are consistent with those derived from [CII] for the QSO host galaxy, we discard this scenario and conclude that this line is due to H$_2$O from the ISM of the QSO host galaxy and not to [CII] from the proximate DLA.   

\begin{figure}
        \centering
        \includegraphics[width=1\linewidth]{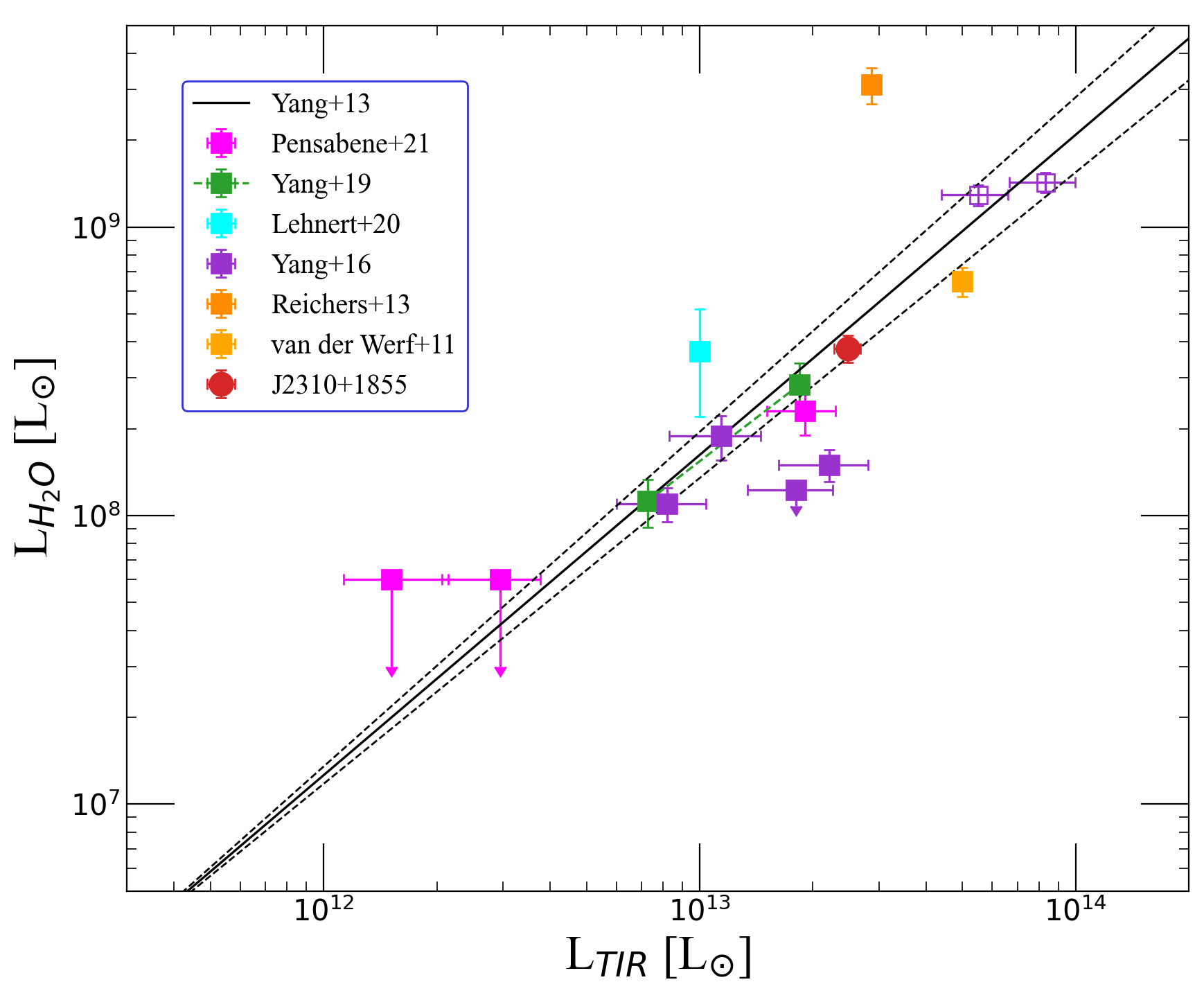}
        \caption{$L_{\rm H_2O}-L_{\rm TIR}$ relation for QSO J2310+1855 and a compilation of high-redshift QSOs and SMGs with H$_2$O detection. The solid black line traces the best power-law fit from \cite{yang2013}. The dashed lines are the $1\sigma$ confidence limit.}
        \label{h2o-plot}
\end{figure}

In Fig. \ref{h2o-plot} we compare the $\rm H_2O\ 3_{22}-3_{13}$ line luminosity with the total IR luminosity of J2310+1855 obtained through the fit of the dust SED (MBB component only, see Sect. \ref{sec:dust}),  together with a compilation of $z\sim 2.5-3.5$ Hy/ULIRGs from \citet{yang2016}, APM 08279+5255 at $z\simeq 3.9$ \citep{vanderwerf2011}, the $z\simeq 6.34$ SMG HFLS 3 \citep{reichers2013}, the QSO host galaxy J0439+1634 at $z\simeq 6.5$ \citep{Yang2019}, QSO BR1202-0725 at $z = 4.695$ \citep{Lehnert2020}, and two QSO host galaxies at $z>6$ PJ231-20 and PJ308-21 \citep{Pensabene2021}, which all have a $\rm H_2O\ 3_{21}-3_{12}$ detection, except for the one of \citet{Lehnert2020}, which has $\rm H_2O\ 2_{20}-2_{11}$. We also show the best power-law fit found by \citet{yang2013} with $1\sigma$ confidence limits, $L_{\rm H_2O} \propto L_{\rm TIR}^{1.1\pm 0.5}$. Our result for J2310 agrees very well with the increasing trend of the data and the linear relation of \citet{yang2013}.  We note that when the contribution of the hot dust heated by the QSO is implied, it would imply $L_{\rm TIR}=8.4\times 10^{13}\rm ~L_\odot$, and would move J2310+1855 well below the correlation.  These data suggest that the  $\rm H_2O$ line is mainly excited by IR-pumping mechanism of the dust-reprocessed UV radiation field  in the ISM of the host galaxy \citep{yang2013, yang2016, Pensabene2021}.

\subsection{Environment of the QSO}
\label{sec:env}

First, we studied the field at the position of the line emitter presented in \cite{dodorico2018}, called Serenity-18. Serenity-18 is a candidate CO(6-5) emitter at a redshift $z=5.9386$, associated with a proximate DLA system located at $z=5.938646 \pm 0.000007$ on the QSO sightline. 
The frequency setup of this observation covers the [CII] emission line, expected at 273.9 GHz, and the sub-mm continuum of Serenity-18 (Fig. \ref{cont_sere}). At the position of Serenity-18 (RA, DEC=[23:10:38.44, 18:55:21.95]), we derived a 3$\sigma$ upper limit on the 265 GHz dust continuum of $S_{265.4 \rm GHz} < 0.026$ mJy. Assuming a dust temperature of $40$ K, an emissivity $\beta=1.5$ \citep{sommovigo2021}, and a source size equal to the clean beam size, this translates into an upper limit on the dust mass of  $M_{\rm dust}< 6\times 10^6\ \rm M_{\odot}$.
For the [CII] emission line, we derived a $3\sigma$ upper limit of $Sdv_{[\rm CII]}< 28.2$ mJy $\rm km\ s^{-1}$, assuming a line width of 200 $\rm km\ s^{-1}$ (equal to the FWHM reported in \citealt{dodorico2018}). This implied an upper limit on the luminosity of the [CII] line of $L_{\rm [CII]}< 2.7\times 10^7 \rm ~L_\odot$. 
Adopting the ${\rm SFR} - L_{[\rm CII]}$ correlation from \citet{carniani2018}, this 
yielded an upper-limit star formation rate of ${\rm SFR}<2.5~ M_\odot yr^{-1}$.  By applying the ${\rm SFR} - L_{[\rm CII]}$ correlation found by \citet{herrera-camus2018} for star-forming main-sequence galaxies with normal star formation efficiency, we found a consistent upper limit of ${\rm SFR}<3\rm ~M_\odot yr^{-1}$.
This upper limit on the SFR, together with the luminosity ratio of CO(6-5) and [CII], $R_{\rm CO-[CII]}>15$, makes it unlikely that DLA J2310+1855 has a [CII]-emitting counterpart, suggesting that the line emitter Serenity-18 is most likely a foreground source at lower redshift, as discussed in \citet{dodorico2018}. 

Another scan of the data cube did not reveal any line emitter at any position down to a luminosity  $L_{[\rm CII]}=  2.7\times 10^7 \rm ~L_\odot$, assuming a typical line width of 200 $\rm km\ s^{-1}$. 
One continuum emitter was detected (Fig. \ref{cont_sere}, Sect. \ref{sec:cII-h2o}). 
We conclude that this QSO is isolated. It does not show close companions or signatures of an ongoing merger. 

\section{Discussion and summary}
\label{sec:summary}
We reported results from a deep ALMA observation of the sub-mm continuum, [CII], and H$_2$O emission lines with $900$ pc resolution, complemented by multiple ALMA archival datasets probing the infrared continuum emission of the $z\sim 6$ QSO J2310+1855. The $900$ pc resolution of this dataset allowed us to perform a detailed study of dust properties and cold gas kinematics and dynamics.

 The accurate sampling of the QSO SED, especially at lower wavelengths ($\lambda\sim 10-100\ \mu$m), allowed us to constrain the dust temperature, $T_{\rm dust} = 71\pm 4$ K, dust mass,  $M_{\rm dust}= (4.4 \pm 0.7) \times 10^8\ \rm M_{\odot}$ , and emissivity index, $\beta = 1.86\pm 0.12$ with high accuracy. We modelled the large-scale dust in the ISM and dusty torus emission with an MBB and dusty torus templates. The values of dust temperature and dust mass are $ \text{about two}$ times higher and $ \text{about four}$ times lower, respectively, than those derived by \citet{shao2019}. One likely cause of discrepancy can be the different treatment of the dusty torus contribution in the SED. While we used a library of SED templates, SKIRTOR, \citet{shao2019} used the cumpy AGN tori in a 3D geometry (CAT 3D) model \citep{honig2017} to represent the near-infrared and mid-infrared contributions from the AGN dust torus. Different modelling of the AGN torus can in principle affect the determination of the parameters related to the large-scale dust emission that is modelled with an MBB. In our analysis, we showed that our prescription for the torus does not influence our results for the MBB, but this may not be the case for the modelling presented in \citet{shao2019}. Moreover, the discrepancy for $M_{\rm dust}$ can be partially explained by the fact that \citet{shao2019} fixed the dust emissivity index, $\beta$, at $1.6$, while we left it as a free parameter, obtaining a value of the emissivity of $\beta=1.86\pm0.12$ (see Table \ref{table-sed-res}). Given that all parameters are strongly correlated, the dust mass increases when $\beta$ decreases at a fixed dust temperature (see Fig. \ref{sed}). We derived an ${\rm SFR} = 1240^{+310}_{-260}\ \rm M_{\odot} yr^{-1}$, accounting for the QSO contribution to dust heating and adopting a Chabrier IMF. Assuming a Salpeter IMF, as in \citet{shao2019},  would imply an SFR higher by factor of 1.7 (i.e. SFR$\sim 2108 \pm 500 ~\rm M_{\odot}yr^{-1}$). We obtained a GDR$= 101\pm 20$, using our estimate of $M_{\rm dust}$, and $M_{\rm H_2} = 4.4 \pm 0.2 \times 10^{10} \rm ~M_{\odot}$ \citep{li2020, feruglio2018}. This GDR is a factor of 5 larger than the one derived by \citet{shao2019}, who reported a very low ${\rm GDR}=26\pm6$. Our value of GDR is consistent with the value normally assumed for high-z QSO (${\rm GDR}=100$, e.g. \citealt{walter2020,wang2019}) and measured for a few of them (${\rm GDR}=100-300$, \citealt{bischetti2021}). 

Our analysis of the [CII] kinematics and dynamics, based on the $^{\rm 3D}$Barolo dynamical model, indicates a disk that is inclined at $i\sim 25$ deg, which is rotationally supported with $v_{\rm rot}/\sigma_{\rm gas} \sim 6$. The gas Toomre parameter is in the range $Q\sim 1-5$ out to r=1.7 kpc, which indicates a marginally unstable disk. The high resolution and high S/N of the [CII] observation allowed us to retrieve the best estimate for the dynamical mass $M_{\rm dyn} = 5.2\times 10^{10}\ \rm M_{\odot}$ within $r = 1.7$ kpc. This enabled us to derive a rough estimate of the stellar mass of the QSO host galaxy, $M_{*}=M_{\rm dyn}-M_{\rm H_2}-M_{\rm BH}\sim 3\times 10^9\ \rm M_{\odot}$, using $M_{\rm BH}=5\times 10^9 \rm  ~M_{\odot}$ derived from the MgII emission line profile (Mazzucchelli in prep). This stellar mass, together with the AGN-corrected SFR, place the QSO host galaxy well above the main sequence for star-forming galaxies at $z\sim 6$ (see e.g. \citealt{mancuso2016, pearson2018}), indicating a strongly star-bursting host galaxy. 

\begin{figure}
        \centering
        \includegraphics[width=1\linewidth]{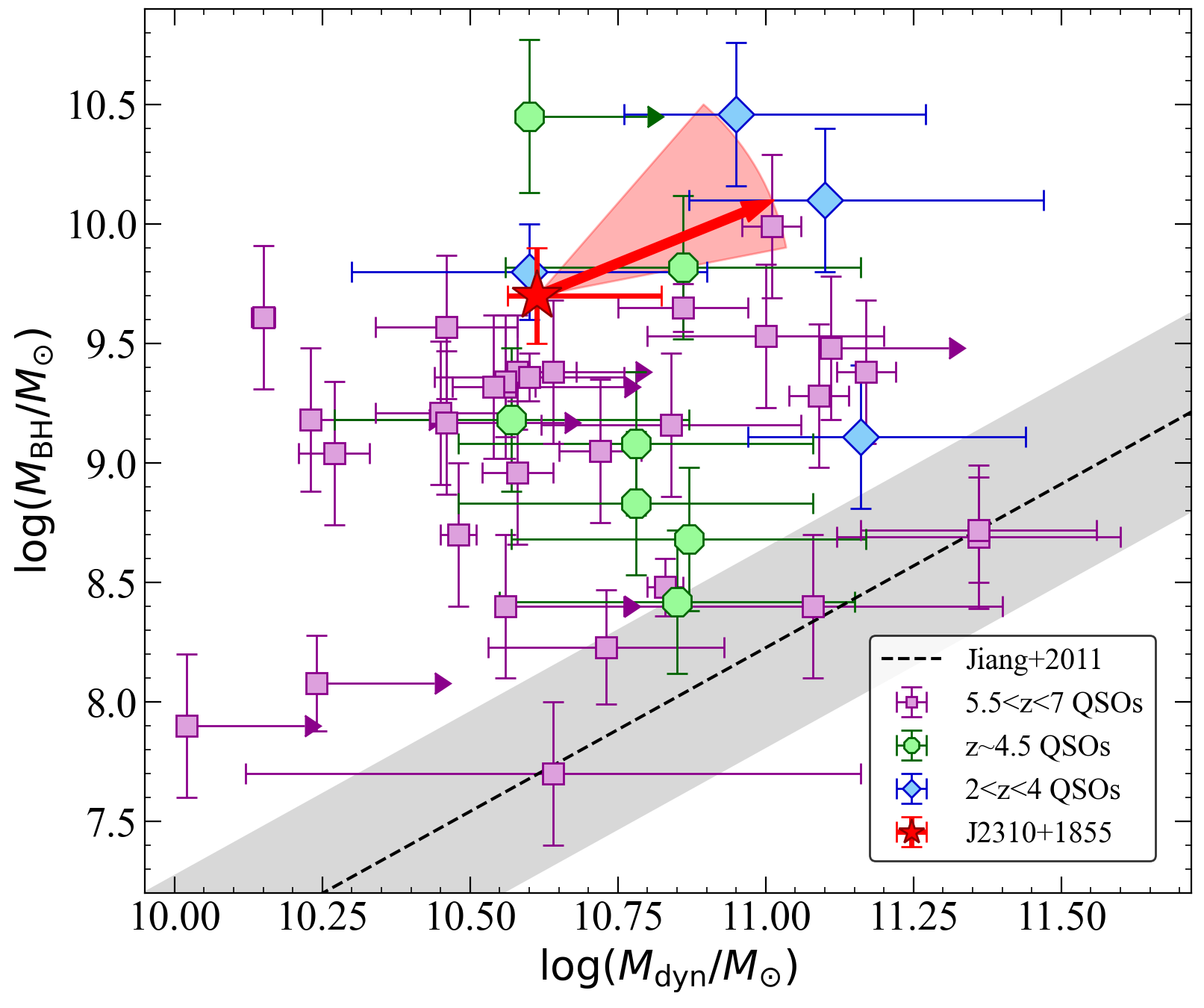}
        \caption{BH mass as a function of the dynamical mass for J2310+1855 (red star), compared with WISSH QSOs at $z\sim 2-4$ (blue diamonds, from \citealt{bischetti2021}), and luminous $z\sim 4-7$ QSOs (green dots and violet squares, from \citealt{venemans2016,venemans2017a,willott2013,willott2015,willott2017,kimball2015,trakhtenbrot2017, feruglio2018, mortlock2011, derosa2014,kashikawa2015, neeleman2021}). \citet{jiang2011} used bulge masses, while all the other dynamical masses refer to the gas disk component. For J2310, the slope of the red arrow, with its uncertainty (shadowed red region), indicates how much the growth efficiency of the SMBH is slowing down with respect to the growth of the host galaxy.}
        \label{mbhmdyn}
\end{figure}

The star formation efficiency of the host galaxy, defined as ${\rm SFR}/M_{\rm H_2}\simeq {\rm SFR}/M_{\rm dyn}$, is indeed $2.4 \times 10^{-8}\rm ~yr^{-1}$, bearing in mind that we used a spatially integrated dust-reprocessed SFR. On the other hand, we derive a BH growth efficiency, \footnote{$\dot M_{\rm BH}=L_{\rm bol}/(\epsilon ~c^2)$, where $\epsilon$ is the radiative efficiency, and $c$ is the speed of light.} $\dot M_{\rm BH}/M_{\rm BH}=1.1\times 10^{-8} \rm ~yr^{-1}$, where we used the BH mass derived from  MgII ($M_{\rm BH}=5\times 10^{9} \rm ~M_{\odot}$, Mazzucchelli in prep.), and assumed $\epsilon=0.1$ (e.g. \citealt{marconi2004, martinez2009, fernandes2015}). 
In Fig. \ref{mbhmdyn} we show the $M_{\rm BH}$ and $M_{\rm dyn}$ for SDSS J2310+1855 and a compilation of QSOs at different redshifts, comparing them with the local $M_{\rm BH}-M_{\rm dyn}$ relation found by \citet{jiang2011}\footnote{We note that \citet{jiang2011} used bulge dynamical masses, while for all the other QSOs, the dynamical mass refers to the gas disk component.}. The majority of  QSOs, including J2310+1855, are above the local relation in the BH dominance regime \citep{volonteri2012}.
For J2310+1855, we found $\dot M_{\rm BH}/M_{\rm BH}<{\rm SFR}/M_{\rm dyn}$, suggesting that AGN feedback is effectively slowing down the accretion onto the SMBH, while the host galaxy is growing fast \citep{volonteri2012}. In particular, the BH growth efficiency is $\sim$50\% lower than that of the QSO host galaxy, as represented by the the slope of the red arrow in Fig. \ref{mbhmdyn} (a slope of $45$ deg corresponds to the case of SFR$/M_{\rm dyn}=\dot M_{\rm BH}/M_{\rm BH}$). The shadowed red region arises from the uncertainties on $M_{\rm BH}$ and $M_{\rm dyn}$. It is important to bear in mind that the slope of the arrow corresponds to the specific case of (1) $M_{\rm dyn}\gtrsim M_{\rm gas}$, which is plausible at high z, (2) an SFR that is computed assuming, conservatively, that the FIR luminosity from SFR is $\text{about half}$ of the total FIR luminosity, and (3) a BH radiation efficiency of 10\%, since a precise determination of this parameter would require a dedicated study of the growth and accretion history of this SMBH, which is beyond the scope of this work. Therefore, a BH growth rate that is lower or similar to that of its host galaxy can be considered a bona fide result. One of the likely causes of the slow-down of the SMBH accretion are radiatively driven AGN winds that impact on the accreting matter, providing enough momentum to stop further accretion, and which can further propagate outwards on the scale of the host galaxy. 
In SDSS J2310+1855, the SMBH accretion may be limited by the ionised wind traced by a C IV broad absorption line (BAL) system with velocity $v_{\rm BAL}=26900$ $\rm km\ s^{-1}$ and balnicity index $BI=600$ \citep{bischetti2022}. 

SDSS J2310+18655 also shows evidence of a [CII] outflow approximately located in the central $\text{}\text{} $kpc, with an outflow mass $M_{\rm out} = 3.5\times 10^8\ \rm M_{\odot}$. This is about $5\%$ of the neutral gas mass in the disk, consistent with expectations of recent zoom-in hydrodynamical simulations presented by \citet{valentini2021}. 
We estimated the mass outflow rate in the range $\dot M_{\rm out}= 1800-4500\rm ~M_\odot yr^{-1}$, which also agrees well with the results of zoom-in cosmological hydrodynamical simulations of the $z\sim 6$ luminous QSO analysed in \citet{barai2018}, who found $\sim 2000-3000\ \rm M_{\odot}yr^{-1}$ within $1$ kpc.
Individual detections of cold gas outflows in very high redshift QSOs are still relatively rare. Currently known cold outflows in $z\sim 6$ QSOs are SDSS J114816.64+525150.3 ($z=6.4$) by \citet{maiolino2012}, whose outflow rate ($\dot M_{\rm out}\gtrsim 3500\ \rm M_{\odot}\ yr^{-1}$) is broadly consistent with our range, while [CII] outflows in QSO HSC J124353.93+010038.5 ($z=7.07$) and HSC J120505.09-000027.9 ($z=6.72$,  \citealt{izumi2021a,izumi2021b}) have lower-limit outflow rates of about $\gtrsim 100\ \rm M_{\odot}\ yr^{-1} $. A stacking analysis of a large sample of $z>4$ QSOs suggests average outflow rates of $\sim 100\ \rm M_{\odot}\ yr^{-1} $ \citep{bischetti2019b}. We estimated $\dot E_{\rm out} \sim 0.0005-0.001\ L_{\rm bol}$ and $ \dot P_{\rm out}/\dot P_{\rm AGN} \sim 0.6-1.4$. Comparing these results with the scaling relations derived by \citet{fiore2017}, we note that $\dot E_{\rm out}/L_{\rm bol}$ is consistent with the scaling for ionised winds, and $\dot P_{\rm out}/ \dot P_{\rm AGN}$  agrees with expectations for momentum-conserving winds.
The BAL is only detected through the C IV absorption trough \citep{bischetti2022}. Its outflow mass and energetics therefore cannot be reliably estimated \citep{borguet2013,byun2022}. This means that current data do not allow us to compare the energetics of these two wind phases. 

Although [CII] probes mildly ionised gas, it has recently been proposed as a possible molecular gas tracer.  \citet{zanella2018} derived a $L_{[\rm CII]}$-to-$\rm H_2$ conversion factor of $\alpha_{\rm [CII]}\sim 22\ \rm M_{\odot}/L_{\odot}$ for star-forming galaxies at $z\sim2$. Applying this conversion to J2310+1855 would imply a molecular mass of $M_{\rm H_2}\sim 1.1\times 10^{11}\ \rm M_{\odot}$. In addition to being a factor of 4 larger than the molecular mass derived from CO \citep{feruglio2018}, this is also $ \text{about three}$ times higher than the dynamical mass derived from [CII]. This suggests that the empirical correlation between [CII] luminosity and molecular mass (e.g. \citet{zanella2018}) does not apply to high-z hyper-luminous QSOs.
We estimated the neutral gas mass of the disk based on [CII] emission, $M_{\rm HI} = 6.6\times10^9\ \rm M_{\odot}$, which is significantly lower than the molecular mass based on CO lines \citep{feruglio2018, shao2019}. The total gas surface density $\Sigma_{\rm (HI + H_2)}(= 13809 \ \rm M_{\odot}\ pc^{-2})$ within the [CII] half-light radius, together with the AGN-corrected SFR surface density $\Sigma_{\rm SFR} = (521\ \rm M_{\odot}yr^{-1}kpc^{-2})$, shows that the host galaxy lies above the region of the local Kennicutt-Schmidt (KS) relation that is usually occupied by starbursting galaxies (e.g. \citealt{bigiel2008}).
Recent zoom-in high-resolution simulations \citep{pallottini2022} and semi-analytical models \citep{vallini2021} found burstiness parameters, $k_{\rm s}$, in the range $\sim 3-100$ for $z\gtrsim6$ galaxies. For our QSO, we find $k_{\rm s}\sim 5$, implying a starbursting host galaxy.

Comparing the surface brightness profiles of the continuum, [CII] and CO(6-5), we found that the dust ($r\sim6.7$ kpc) is more extended than [CII] ($r\sim5$ kpc) and CO ($r\sim4.7$ kpc), whereas it is more peaked at the centre. 
A similar behaviour with a steeper dust continuum distribution is seen in other high z QSOs (e.g.  \citet{walter2022}) and has been attributed to the contribution of the QSO to the dust heating.
The ratio of integrated [CII] to TIR luminosity over the whole source is $L_{\rm [CII]}/L_{\rm TIR}\sim 6\times 10^{-5}$.  This [CII] deficit is also predicted for high-z galaxies by semi-analytical models of galaxy evolution (e.g.  \citet{lagache2018}), where  the [CII] deficit arises from the high intensity of the interstellar radiation field. Our estimate of $L_{\rm [CII]}/L_{\rm TIR}$  agrees well with their results at $z\sim 6$ when we extrapolate their predictions at higher $L_{\rm TIR}$. \citet{carniani2018} studied the $L_{\rm [CII]}-{\rm SFR}$ relation for high-z galaxies and reported that the local relation for star-forming galaxies (see \citealt{delooze2014}) is still valid at high z, but with a $ \text{twice higher}$ dispersion than observed locally. Our results agree very well ($<1\sigma$) with the correlation of \citet{carniani2018}, with the relation for local star-forming galaxy of \citet{herrera-camus2018}, and with the results for high-z galaxies of \citet{lagache2018}.

For the first time, we were able to map a spatially resolved H$_2$O v=0 $3_{(2,2)}-3_{(1,3)}$ emission line at $\nu_{\rm obs} = 274.074$ GHz at a statistical significance of $10\sigma$ . Its emission is consistent with a water vapour disk whose kinematics agrees with the [CII] disk. From the observed H$_2$O velocity gradient and adopting $i=25$ deg, we estimated $M_{\rm dyn, H_2O} = 6.4\times10^{10}\ \rm M_{\odot}$ within a diameter of $D = 1.5$ kpc. The luminosity ratio $L_{\rm H_2O}/L_{\rm TIR, MBB}= 1.4\times 10^{-5}$ is consistent with line excitation by dust-reprocessed star formation in the ISM of the host galaxy. However, the faintness of this emission line makes it unsuitable for more detailed dynamical studies. 

Finally, we studied the environment of J2310+1855, scanning the data cube for line emitters. No line emitter was detected down to a $3\sigma$ upper limit of  $L_{\rm [CII]} < 2.7 \times 10^7\ \rm L_{\odot}$, or ${\rm SFR}<2.5\ \rm M_{\odot} yr^{-1}$. We also note that the proximate DLA J2310+1855 did not show any line-emitting counterpart down to this limit, and therefore, the line emitter reported in \citet{dodorico2018}, called Serenity-18, is most likely a lower-z interloper. In the continuum data, we found a low-significance ($4 \sigma$) continuum emitter located 4 arcsec offset from the QSO position, whose physical association with the QSO remains to be confirmed. This led us to conclude that the QSO J2310 does not show any evidence of companions, interaction, or merger at least on scales of  $\sim 50$ kpc. Other observations suggested that 20-50\% of QSOs show mergers or close companions, independent of their luminosity \citep{decarli2018, venemans2020, neeleman2021}.  
 In a recent cosmological simulation of a $z\sim 6$ QSO, \citet{zana2022} computed the number of companions associated with a QSO. Based on their results, we would expect to detect three companions with $L_{\rm [CII]}\sim 10^8 \rm \ L_{\odot}$ within $250$ kpc from the QSO, and seven companions with $L_{[\rm CII]}=2.7\times 10^7 \rm ~L_\odot$ in approximately the same region.

\section{Conclusions}
\label{sec:concl}

 The picture that finally arises is that of an isolated QSO, without evidence of ongoing mergers, that is characterised by a rotationally supported disk with a Toomre parameter $Q_{\rm gas}\sim3$ out to a radius of $1.5$ kpc. The gas kinematics shows evidence of a gaseous outflow within the central kpc, as also supported by the flat rotation curve and the rather high velocity dispersion at the nucleus. To better constrain the nuclear gas kinematics and spatially resolve the outflow, observations with a resolution of $\sim 0.03$ arcsec are required. Moreover, the fact that ${\rm SFR}/M_{\rm dyn}>\dot M_{\rm BH}/M_{\rm BH}$ suggests that the SMBH accretion is slowing down in this QSO, probably owing to the BAL wind seen in CIV, while the stellar mass assembly takes place vigorously in the host galaxy. Our study may suggest that this $z\sim6$ QSO is witnessing the fall of the black-hole dominance phase. In order to test whether this conclusion can be generalised to the entire population of $z\gtrsim 6$ QSOs, we aim to complement this study through the analysis of other QSOs at high z in order to confirm or rule out a particular evolutionary scenario. In particular, to allow precise constraints on the BH and host-galaxy evolutionary paths for a larger sample of high-z QSOs, it is essential to obtain a reliable and accurate determination of the SFR in the host galaxies. The uncertainty on the SFR strongly depends on the accuracy in the estimates of $T_{\rm dust}$ and $M_{\rm dust}$, based on the SED analysis. Therefore, we highlight the need for observations with ALMA bands 8-10 to probe near the peak of the cold dust SED, enabling us to achieve high precision in the determination of the SFR.

\begin{acknowledgements}
      We thank the anonymous referee for her or his careful review of the paper and insightful suggestions. This paper makes use of the following ALMA data: ADS/JAO.ALMA\#2019.1.00661.S, ADS/JAO.ALMA\#2019.1.01721.S, ADS/JAO.ALMA\#2015.1.00584.S, ADS/JAO.ALMA\#2015.1.01265.S, ADS/JAO.ALMA\#2018.1.00597.S, ADS/JAO.ALMA\#2013.1.00462.S, ADS/JAO.ALMA\#2017.1.01195.S. ALMA is a partnership of ESO (representing its member states), NFS (USA) and NINS (Japan), together with NRC (Canada), MOST and ASIAA (Taiwan) and KASI (Republic of Korea), in cooperation with the Republic of Chile. The Joint ALMA Observatory is operated by ESO, AUI/NRAO and NAOJ. RT acknowledges financial support from the University of Trieste. Authors acknowledge support from PRIN MIUR project “Black Hole winds and the Baryon Life Cycle of Galaxies: the stone-guest at the galaxy evolution supper”, contract \#2017PH3WAT. RM acknowledges ERC Advanced Grant 695671 QUENCH, and support from the UK Science and Technology Facilities Council (STFC). RM also acknowledges funding from a research professorship from the Royal Society. SCa, AP and LV acknowledge support from the ERC Advanced Grant INTERSTELLAR H2020/740120 (PI: Ferrara). This paper makes extensive use of \textit{python} packages, libraries and routines, such as \texttt{numpy}, \texttt{scipy} and \texttt{astropy}.
      \textit{Facilities:} ALMA, Herschel. \textit{Software:} CASA (v5.1.1-5, \citealt{mcmullin2007}). 
      
\end{acknowledgements}

%
%

\bibliographystyle{aa} 
\bibliography{biblio}

\end{document}